\documentclass[letterpaper,doc,12pt,floatsintext,longtable]{apa6}

\usepackage[american]{babel}
\usepackage[utf8]{inputenc}
\usepackage{amsmath}
\usepackage{graphicx}
\usepackage{xcolor, colortbl}
\usepackage{caption}
\usepackage{setspace}
\usepackage{threeparttable}
\usepackage{endnotes}
\usepackage{xspace}
\usepackage{afterpage}
\usepackage{bm}
\usepackage{amssymb}
\usepackage{lscape}
\usepackage[colorlinks=true, allcolors=black]{hyperref}
\usepackage[normalem]{ulem}
\usepackage{csquotes}
\usepackage{tikz}
\usepackage{dsfont}
\usepackage{booktabs}
\usepackage{multirow}
\usepackage{tabularx}

\usepackage{setspace}

\usetikzlibrary{decorations.pathreplacing}
\graphicspath{ {} }

\usepackage[style=apa,sortcites=true,sorting=nyt,backend=biber]{biblatex}
\DeclareLanguageMapping{american}{american-apa}
\usepackage{array}
\newcommand{\PreserveBackslash}[1]{\let\temp=\\#1\let\\=\temp}
\newcolumntype{C}[1]{>{\PreserveBackslash\centering}p{#1}}
\newcolumntype{L}[1]{>{\PreserveBackslash\raggedright}p{#1}}
\newcolumntype{R}[1]{>{\PreserveBackslash\raggedleft}p{#1}}

\newcommand{\indep}{\perp\!\!\! \perp}

\usepackage[noend]{algpseudocode}
\usepackage{algorithm}
\usepackage{listings}

\usepackage{arydshln}

\usepackage{tikz}
\usetikzlibrary{matrix,positioning,calc}

\usepackage{subcaption}
\usepackage{amsmath,amssymb,xcolor}
\usepackage{pgf,tikz}
\usetikzlibrary{
    arrows,
    shapes.arrows,
    shapes.geometric,
    shapes.multipart,
    decorations.pathmorphing,
    positioning,
    shapes,
    decorations,
    arrows.meta,
    fit,
    calc
}

\title{\Large Generative AI-Based Monte Carlo Simulation \\[0.35em] 
for Method Evaluation Using Synthetic Multilevel Data}

\shorttitle{AI-Based Monte Carlo Simulation}

\author{\fontsize{13.5pt}{13pt}\selectfont Youmi Suk$^*$, Chenguang Pan, and Weixuan Xiao \vspace{0.3em}
}

\affiliation{Teachers College, Columbia University  \\[0.3em] 
\{ysuk, cp3280, wx2299\}@tc.columbia.edu
\\[0.5em] 
May 7, 2026}

\begin{document}

\maketitle

\begingroup
\renewcommand{\thefootnote}{\fnsymbol{footnote}}
\footnotetext[1]{Corresponding author: Youmi Suk. Email: ysuk@tc.columbia.edu.}
\endgroup

\vspace{-1em}
\begin{center}
\textbf{Abstract}
\end{center}

\noindent
\begin{minipage}{\textwidth}
The role of AI-generated synthetic data has recently been expanded to support realistic Monte Carlo simulations. However, guidance is limited on generating data with multilevel structures and designing simulations based on such data. This study proposes a general framework for AI-based simulation studies to evaluate the predictive performance and parameter recovery of quantitative methods, specifically using multilevel data commonly observed in the social sciences. Our proposed six-stage workflow consists of (i) specifying a method and real data, (ii) training Generative AI with real data, (iii) assessing synthetic data quality, (iv) designing and conducting simulations, (v) evaluating method performance, and (vi) checking robustness. To enhance fidelity in multilevel data generation, we also introduce targeted modifications to diffusion models and Generative Adversarial Networks (GANs). Furthermore, we develop a systematic quality evaluation framework that assesses both within-table and between-table fidelity, and discuss how AI-based simulation designs should differ depending on whether the simulation's objective is predictive performance or parameter recovery. Finally, using empirical multilevel data and multilevel modeling methods, we demonstrate the utility of the proposed AI-based simulation framework. This approach leads to more accurate and honest evaluations of quantitative methods in the real world, unlike traditional simulation studies based on arbitrary simulated scenarios.

\vspace{0.75em}

\end{minipage}

\vspace{1.5em}

\setcounter{secnumdepth}{3}

\section{Introduction}

\subsection{Motivations: Simulations with Realistic Data}

AI-generated synthetic data are increasingly adopted across various domains, including computer vision, healthcare, finance, and education \parencite[e.g.,][]{lu2025machine, suk2025using, assefa2020generating, movahedi2023evaluating}. Modern generative AI (GenAI) methods create synthetic data with minimal or no human intervention in the generation pipeline. Notable examples include Variational Autoencoders \parencite[VAEs;][]{kingma2014auto}, Generative Adversarial Networks \parencite[GANs;][]{goodfellow2014generative}, diffusion models \parencite{ho2020modeling}, and large language models \parencite[LLMs;][]{brown2020language}. Such GenAI methods learn patterns from large datasets and use that knowledge to generate novel outputs that are statistically similar to, but not identical to, their training data \parencite{lu2025machine}. More recently, realistic synthetic data have been utilized in the design of Monte Carlo simulations to numerically evaluate the performance of quantitative methods in real-world scenarios \parencite{suk2025using, athey2024using}. However, existing studies focus on a single table\footnote{A table or tabular data consists of a structured dataset, where columns represent variables and rows represent observations.} or a single objective of simulations (e.g., predictive performance), thereby limiting their general applicability to quantitative methods in the social sciences. Therefore, the overarching goal of this paper is to propose a general AI-based simulation approach by incorporating two main simulation objectives---\emph{predictive performance} and \emph{parameter recovery}---and synthesizing multilevel (multi-relational) data beyond single-level structures.    

Monte Carlo simulations are widely used in statistics, quantitative methodology, and computational sciences to examine a method’s finite-sample properties, assess its robustness to potential violations of underlying assumptions, or evaluate methods where statistical theories are weak or absent \parencite{metropolis1949monte, thomopoulos2013eseential, fan2002monte}. The objectives of simulations can be categorized into two main types: predictive performance and parameter recovery. A predictive performance study evaluates how accurately a model or method can make predictions or forecasts on new data beyond the data used for model training. In contrast, a parameter recovery study assesses how well a statistical method can recover the parameters of a target population of interest, such as a regression coefficient in a random-effects model, to understand the accuracy and reliability of the method.\footnote{A third objective can be a causal effect study, which assesses how accurately a method can estimate treatment effects. Unlike the two main types, such studies must consider causal relationships among variables in data-generating processes, which is beyond the scope of this paper.} Regardless of the specific objective, researchers retain full discretion over data-generating processes in traditional Monte Carlo simulations. They commonly use oversimplified distributions characterized by a high degree of smoothness (e.g., normal or uniform distributions), limited interdependencies among variables, and linear relationships between variables. These distributions may be selected to favor the researchers' specific methods of interest. Under these ``idealized'' settings, the performance of existing and newly developed quantitative methods is not always indicative of their ``true'' performance in the real world \parencite{athey2024using, suk2025using, advani2019mostly}.

However, GenAI methods excel at producing realistic data (e.g., text, image, video, and tabular data) that humans or automatic detectors struggle to distinguish from real data. In particular, specialized GenAI models have been developed to generate synthetic single-table data by addressing common challenges in generation processes, such as mixed data types, inter-variable relationships, and non-Gaussian distributions. Examples include  Tabular VAE \parencite[TVAE;][]{xu2019modeling}, TableGAN \parencite{park2018data}, and Conditional Tabular GAN \parencite[CTGAN;][]{xu2019modeling}. Among these, CTGANs is one of the most widely used GAN-based models for single-table synthesis and uses conditional GANs to handle complex tabular distributions. In contrast, multi-table synthesis is an emerging area, where GenAI methods are tailored to generate multiple interconnected tables, such as Cluster Latent Variable guided Denoising Diffusion Probabilistic Models \parencite[ClavaDDPM;][]{pang2024clavaddpm}. Unfortunately, all existing methods exhibit limited performance in capturing the complex data structures found in educational datasets, specifically those with multilevel structures. For example, CTGAN inherently lacks the ability to generate multilevel data, and ClavaDDPM fails to preserve the original clustering effects because it utilizes latent groups that combine distinct clusters. Therefore, it is still an open question as to how these GenAI models can be modified to generate more realistic multilevel data and effectively integrated into AI-based simulation studies with different objectives.

The main goal of this paper is to propose an AI-based Monte Carlo simulation approach to evaluate the predictive performance or parameter recovery of quantitative methods in the real world, specifically using synthetic multilevel data generated by GenAI. Our approach consists of six stages: (i) specify a statistical method and the scheme of real data, (ii) train GenAI models with real data, (iii) assess synthetic data quality, (iv) design and conduct simulations, (v) evaluate method performance, and (vi) check robustness. We demonstrate the proposed approach using example methods from \Textcite{afshartous2005prediction} for predictive performance and those from \Textcite{huang2017multilevel} for parameter recovery, and we also detail how AI-based simulation designs differ depending on the specific objective.

\subsection{Our Contributions}

We introduce a general AI-based simulation framework tailored for both predictive performance and parameter recovery studies. Extending the work of \textcite{suk2025using},  this framework comprises five key stages, along with an optional sixth stage for robustness checks. Unlike the prior work, in the first stage, we emphasize specifying target quantitative methods and data schema, and in the fourth stage, we employ new generation strategies tailored to those objectives. Specifically, the design of AI-based simulations for predictive performance is straightforward; we generate all variables from synthetic tabular data at once via GenAI models and utilize them directly as a single replication in the simulation. However, for parameter recovery studies, we leverage AI to generate realistic covariates, and we introduce additional nuisance variables within a known data-generating mechanism for the outcome variable to establish the ground truth parameters. See Sections \ref{sec:demo1} and \ref{sec:demo2} for more details. We believe this proposed framework will be widely applicable for evaluating the real-world performance of quantitative methods in the social sciences. 

Our AI-based simulation approach also provides a flexible and efficient tool for generating realistic synthetic data of any size to evaluate method performance in real-world contexts. However, traditional methods relying entirely on researcher-specified data-generating processes struggle to generate data that closely resemble real data distributions, whereas existing semi-synthetic simulations with real data suffer from sample size constraints and restricted variation across synthetic datasets; see Section \ref{sec:review_trad} for more details. In contrast, GenAI methods demonstrate superior capabilities in producing new samples that are statistically similar to the original real data while allowing for flexible control over sample size. Therefore, we believe the proposed AI-based simulation approach establishes a new standard for the evaluation of quantitative methods in the real world.

Moreover, we provide modifications to existing GenAI models for producing more realistic multilevel data. For example, ClavaDDPM is not capable of recovering an intra-class correlation (ICC) close to that of real data because it uses re-labeled clusters rather than original intact clusters. While using re-labeled clusters offers certain benefits, it results in a near-zero ICC. Since the ICC is an important simulation factor in multilevel studies, we propose a variable decomposition strategy to preserve the ICC value observed in real data. This transforms individual-level variables into cluster-demeaned and cluster-mean components, and we use these transformed variables during the training stage. 
Regarding CTGAN, we propose a new post-processing procedure that enforces within-cluster coherence of cluster-level variables while maintaining the similarity between synthetic and real data. See Section \ref{sec:AI_multilevel} for more details of the modifications.   

Finally, we propose a formal framework for evaluating the quality of synthetic multilevel data with respect to within-table fidelity, between-table fidelity, and machine learning (ML) efficacy. While most existing quality metrics are designed for single-table data, there are few metrics for multilevel data, like referential integrity and $k$-hop correlation \parencite{sdmetrics, pang2024clavaddpm}. We summarize existing quality metrics for each domain and introduce new metrics to evaluate quality. For example, we draw on the rich literature on multilevel modeling \parencite[e.g.,][]{raudenbush2002hierarchical, snijders2012multilevel} to develop new quality metrics for between-table fidelity: the \emph{ICC similarity} between synthetic and real data, which measures the similarity of individual outcomes within clusters compared to those in other clusters, and the \emph{reliability similarity} of the cluster mean outcomes. See Section \ref{sec:qual_eval} for more details on quality evaluation.

The remainder of this paper is organized as follows. Section \ref{sec:review} reviews traditional methods for data generation and GenAI models for tabular data. Section \ref{sec:AIsimu} introduces our AI-based simulation approach for evaluating quantitative methods. Sections \ref{sec:demo1} and \ref{sec:demo2} demonstrate the proposed approach for predictive performance and parameter recovery studies, respectively. Section \ref{sec:alation} presents the results of the robustness checks, and Section \ref{sec:con} concludes with a discussion of our findings and future directions. \raggedbottom

\section{Review}\label{sec:review}

\subsection{Traditional Methods for Data Generation}\label{sec:review_trad}

Many traditional methods for generating data frequently use researcher-defined, often unrealistic scenarios \parencite[e.g.,][]{morris2019using, fan2002monte, huang2017multilevel, afshartous2005prediction}. In these cases, the marginal distributions of variables are typically assumed to follow smooth parametric forms, such as normal or uniform distributions. These scenarios also tend to exhibit limited dependencies between variables, both within the same level and across different structural levels (e.g., student, school, neighborhood). When dependencies are modeled, they usually take a simple linear form and are often generated using multivariate normal distributions. Furthermore, the diverse measurement scales (e.g., nominal, ordinal, numerical) found in real data are frequently ignored. As a result, these data-generating processes are largely determined by the researcher's discretion or preference and deviate from real data distributions. 

Alternatively, researchers can generate new data by sampling from an estimated distribution derived from real data. These methods first determine statistical parameters (e.g., the mean, variance, and covariance) of the real data distribution using techniques like maximum likelihood estimation (MLE) or kernel density estimation (KDE), and then sample synthetic observations from these estimated distributions \parencite{bishop2006pattern, hastie2009elements, athey2024using}. While there are advantages of each technique, they have their own limitations, such as parametric modeling assumptions in MLE methods and over-smoothing issues in KDE methods. 

Moreover, researchers can use more realistic data via semi-synthetic designs \parencite[e.g.,][]{hill2011bayesian} or empirical Monte Carlo studies \parencite[e.g.,][]{advani2019mostly}. Unlike the purely simulated data above, these approaches anchor the generation process to real data so that the synthetic data inherit as much of its empirical structure as possible. For example, semi-synthetic data are often generated by adding normally distributed errors to outcomes modeled on real data. While these approaches reflect the empirical distributions to some extent, they rely on potentially unrealistic assumptions (e.g., additive Gaussian errors) and are limited in scalability (e.g., fixed sample size) and flexibility for introducing controlled variation in the generated data.

\subsection{Generative AI for Tabular Data}

Unlike traditional statistical methods, GenAI models are based on deep learning and automatically learn characteristics from real data to produce synthetic data that statistically resembles the original input. Over the past decade, these models have been utilized to generate synthetic tabular data by handling common challenges in the data, such as mixed data types and complex data distributions \parencite{shi2025comprehensive}.

First, VAE-based methods adapt the encoder-decoder architecture to handle mixed data types and inter-variable dependencies. An encoder network compresses the input data into a probabilistic latent space (typically a Gaussian distribution), while a decoder network samples from this latent space to reconstruct the original input data. The model is trained to maximize the Evidence Lower Bound \parencite[ELBO;][]{kingma2014auto},  which simultaneously optimizes accurate data reconstruction and regularizes the latent space to be smooth and continuous.  Examples include TVAE \parencite{xu2019modeling} and VAEM \parencite{Ma2020VAEM}. While these VAE-based methods demonstrate stable training and meaningful latent space representations, they often produce overly smooth outputs. 

Second, GAN-based methods involve a min-max game between two competing neural networks \parencite{goodfellow2014generative}. A generator creates synthetic data from a random noise input, while a discriminator acts as an evaluator to distinguish between real data and synthetic instances produced by the generator. The generator's goal is to produce samples realistic enough to fool the discriminator, and the discriminator's goal is to improve its ability to identify synthetic samples. GAN-based methods produce high-fidelity outputs and can generate conditional outputs (i.e., incorporating a conditional vector), for example via conditional GANs \parencite[cGANs;][]{Mirza2014ConditionalGAN}. Although earlier versions of GAN-based methods suffered from training instability, mode collapse, and poor multimodal representation, more advanced models, like CTAB-GAN \parencite{zhao2021ctabgan} and CTGAN \parencite{xu2019modeling}, have mitigated these issues. 

Third, diffusion models are a newer class of models distinct from VAEs and GANs, and consist of two phases \parencite{ho2020modeling}. A noising (forward) process gradually adds random noise (often Gaussian) to a data sample over many small steps until the sample becomes near pure noise. A denoising (reverse) process then reverses this noising by gradually removing the noise step-by-step to generate new, realistic samples. Diffusion models offer stable training and generate high-fidelity, diverse outputs, often outperforming GANs. However, they show slow sampling speeds and high computational costs. Examples include TabDDPM for single tables \parencite{Kotelnikov2023TabDDPM} and ClavaDDPM for multi-tables \parencite{pang2024clavaddpm}. 

Lastly, LLMs are transformer-based models trained on vast amounts of text data. They can be adapted for tabular data generation by framing the task as a text generation problem. These models can employ prompt-based methods that rely on in-context learning capabilities, using carefully designed prompts without parameter updates or fine-tuning. Or they can employ fine-tuning methods, which update the parameters of a pre-trained LLM to adapt it to specific tabular datasets. Although LLM-based methods demonstrate excellent semantic understanding and in-context learning, they are subject to hallucinations, inconsistent formatting, and computationally expensive to train \parencite{shi2025comprehensive}. Consequently, given these trade-offs, we utilize GANs and diffusion models to generate multilevel tabular data for AI-based simulation studies in this paper.  

\section{AI-Based Simulation Approach for Method Evaluation}\label{sec:AIsimu}

In this section, we introduce our AI-based simulation framework and provide more detailed explanations of the first three stages: input data specification, model training, and quality evaluation.  

\subsection{AI-Based Simulation Workflow}\label{sec:AIsimu_Proce}

\begin{figure}[!ht]
    \centering
    \includegraphics[width=0.95\linewidth]{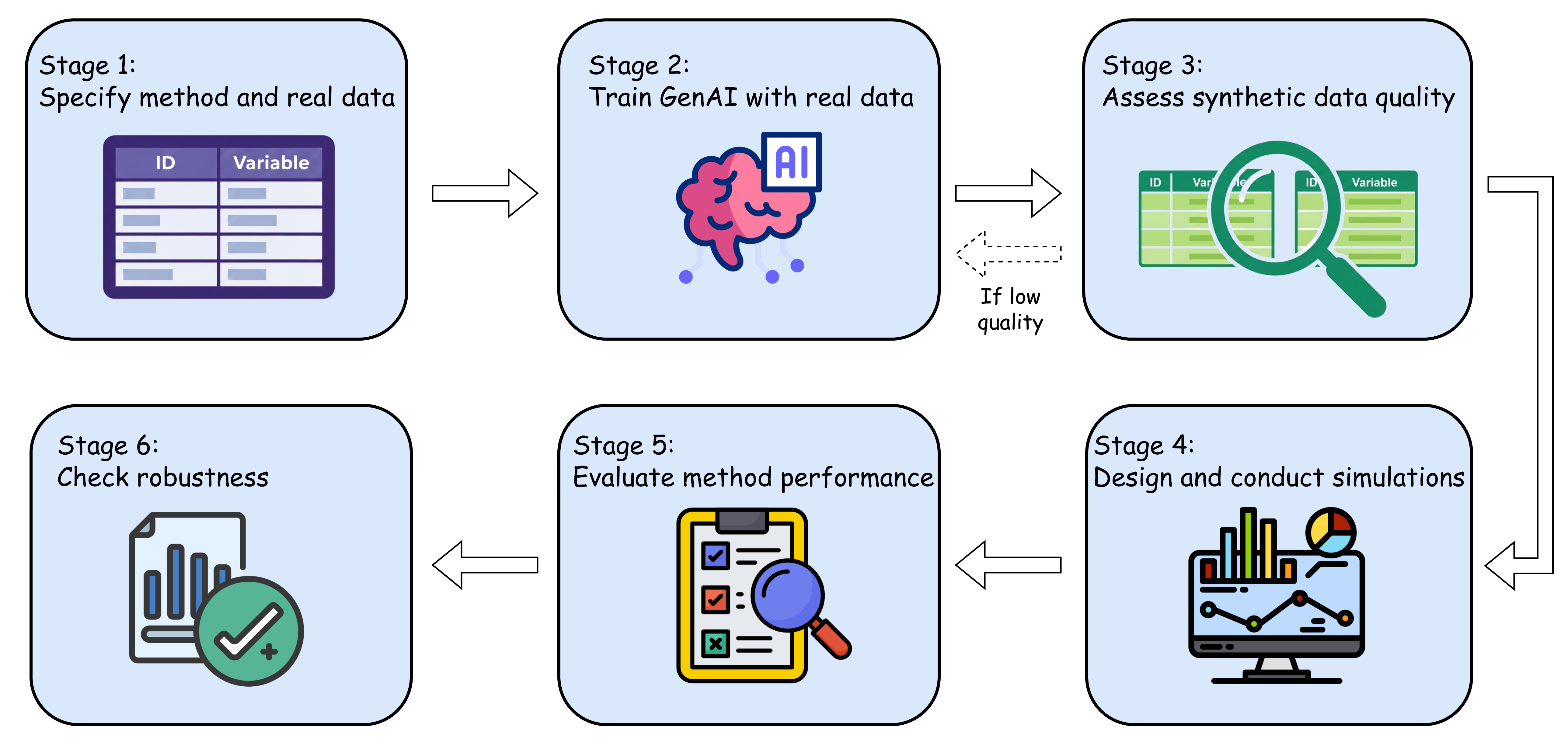}
    \caption{Procedure of the proposed AI-based simulation approach}
    \label{fig:AIsimu}
\end{figure}

Our AI-based simulation approach consists of five core stages, along with an optional sixth stage for robustness checks. See Figure \ref{fig:AIsimu} for an illustration.  

\textit{Stage 1: Specify a quantitative method and the schema of real data.} Researchers first select quantitative method(s) to be evaluated, and specify the schema (i.e., metadata) of the real data. This includes data types (e.g., numerical, categorical) and structural relationships (e.g., primary key and foreign key for multilevel structures). Accurate schema specification is crucial for enabling the AI to generate high-quality synthetic datasets.

\textit{Stage 2: Train GenAI models with real data.} Researchers select GenAI models based on the data types and structural properties specified in Stage 1. These models are then trained using real data, following the standard AI model development protocols \parencite[e.g.,][]{goodfellow2016deep}.

\textit{Stage 3: Assess the quality of synthetic data.} Researchers determine the desired sample size (e.g., the number of clusters, the total sample size) to generate from the trained GenAI model in Stage 2, which typically defaults to the size of the original input data. The similarity between the synthetic data and the real data is then assessed using both visual inspection and formal metrics. Specifically, quality evaluation for multilevel synthetic data focuses on within-table fidelity, between-table fidelity, and ML efficacy; see Section \ref{sec:qual_eval} for details. If synthetic datasets are of insufficient quality, researchers need to adjust the hyperparameters (e.g., epochs, learning rates) of the selected GenAI model and iterate this process until high-quality data are produced. The final output of this stage is a fine-tuned GenAI model (e.g., saved as a Python pickle file, .pkl) that can produce datasets that have passed the quality check.

\textit{Stage 4: Design and conduct simulations.} Using the fine-tuned GenAI model, researchers design and execute their simulation study. Key design issues include selecting simulation factors, defining performance metrics, determining the number of replications, and setting target parameters, similar to traditional simulation studies. Researchers can tailor the simulation design to generate datasets that systematically vary according to a posited data-generating model, in order to examine controlled variations (e.g., intentionally forcing zero correlation between certain variables) and their impact on method performance. Once the procedure is outlined, simulation jobs are executed using computational tools, such as R, Python, or SAS. See Sections \ref{sec:demo1} and \ref{sec:demo2} for two specific demonstrations: one for prediction performance and the other for parameter recovery. 

\textit{Stage 5: Evaluate method performance.} Researchers assess the performance of the chosen quantitative method(s) based on the simulation results and the metrics defined in Stage 4. If applicable, researchers may compare AI-based simulation results with those of traditional simulations based on arbitrary simulated scenarios.

\textit{Stage 6: Check robustness.} While optional, this stage assesses whether modifying the specific components of the generation process (e.g., from optimal to sub-optimal settings) influences synthetic data quality and simulation outcomes, and determines whether the conclusions drawn in Stage 5 remain robust.

\subsection{Multilevel Data: Multiple Tables and Join-as-One Table}

In the first stage of our proposed AI-based simulation approach, researchers should select a target dataset and define its schema. This paper focuses on multilevel data to evaluate multilevel modeling methods using the proposed AI-based simulation approach. Figure \ref{fig:muldata} illustrates multilevel data in two distinct formats, where students are nested within schools. The multiple-table format consists of a parent table (e.g., the school table) and a child table (e.g., the student table). The student table contains two key variables: \texttt{Student\_ID} and \texttt{School\_ID}. \texttt{Student\_ID} serves as the primary key, while \texttt{School\_ID} serves as a foreign key that links student records to the school table. Each variable in these tables is specified by its data type (e.g., numerical, categorical). These variables can be further specified by their data formats (e.g., integer, float) and the range of data (e.g., min, max).

\begin{figure}[!h]
\centering
  \subcaptionbox{Multiple tables \label{fig:multiple}}[1\textwidth]
  {
    \begin{tikzpicture}[>=latex,scale=0.85, every node/.style={transform shape}]

    \matrix (school) [
        matrix of nodes,
        nodes in empty cells,
        nodes={font=\footnotesize, minimum height=4mm, inner ysep=0.5pt},
        column sep=-\pgflinewidth,
        row sep=-\pgflinewidth,
        column 1/.style={anchor=west,  minimum width=3.2cm},
        column 2/.style={anchor=west, minimum width=3.0cm},
        ampersand replacement=\&
    ]{
      School\_ID      \& Primary Key \\
      SchoolRegion    \& Categorical \\
      SchoolLocation  \& Categorical \\
      SchoolClimate   \& Numerical   \\
      \dots           \& \dots       \\
    };

    \node[anchor=south west] at ($(school-1-1.north west)+(0,2mm)$) {\textbf{School Table}};

    \draw (school-1-1.north west) rectangle (school-5-2.south east);
    \draw (school-1-1.north east) -- (school-5-1.south east);
    \draw (school-1-1.south west) -- (school-1-2.south east);

    \matrix (student) [
        matrix of nodes,
        nodes in empty cells,
        nodes={font=\footnotesize, minimum height=4mm, inner ysep=0.5pt},
        column sep=-\pgflinewidth,
        row sep=-\pgflinewidth,
        right=3cm of school,
        column 1/.style={anchor=west,  minimum width=4.2cm},
        column 2/.style={anchor=west, minimum width=3.2cm},
        ampersand replacement=\&
    ]{
      Student\_ID      \& Primary Key \\
      School\_ID       \& Foreign Key \\
      Age             \& Numerical   \\
      Female          \& Categorical \\
      FamilyType      \& Categorical \\
      SocioeconomicStatus \& Numerical   \\
      \dots           \& \dots       \\
    };

    \node[anchor=south west] at ($(student-1-1.north west)+(0,2mm)$) {\textbf{Student Table}};

    \draw (student-1-1.north west) rectangle (student-7-2.south east);
    \draw (student-1-1.north east) -- (student-7-1.south east);
    \draw (student-2-1.south west) -- (student-2-2.south east);

    \draw[<->] (school-1-2.east) -- (student-2-1.west); 

    \end{tikzpicture}
  }
  
  \vspace{0.5cm} 

  \subcaptionbox{Join-as-one table \label{fig:single}}[1\textwidth]{
    \begin{tikzpicture}[>=latex,scale=0.85, every node/.style={transform shape}]

    \matrix (student) [
        matrix of nodes,
        nodes in empty cells,
        nodes={font=\footnotesize, minimum height=4mm, inner ysep=0.5pt},
        column sep=-\pgflinewidth,
        row sep=-\pgflinewidth,
        column 1/.style={anchor=west,  minimum width=4.2cm},
        column 2/.style={anchor=west, minimum width=3.2cm},
        ampersand replacement=\&
    ]{
      Student\_ID      \& Primary Key \\
      Age             \& Numerical   \\
      Female          \& Categorical \\
      FamilyType      \& Categorical \\
      SocioeconomicStatus \& Numerical   \\
      School\_ID       \& Categorical \\  
      SchoolRegion    \& Categorical \\
      SchoolLocation  \& Categorical \\
      SchoolClimate   \& Numerical   \\
      \dots           \& \dots       \\
    };

    \node[anchor=south west] at ($(student-1-1.north west)+(0,2mm)$) {\textbf{Merged Table}};

    \draw (student-1-1.north west) rectangle (student-10-2.south east);
    \draw (student-1-1.north east) -- (student-10-1.south east);
    \draw (student-1-1.south west) -- (student-1-2.south east);

    \end{tikzpicture}
  }
  \caption{The schema of multiple tables (a) and join-as-one table (b) for multilevel data.}
  \label{fig:muldata}
\end{figure}

In contrast, the join-as-one table format merges multiple tables into a single table, where \texttt{Student\_ID} variable serves as the primary key. All other variables are classified as either numerical or categorical, with \texttt{School\_ID} treated as a categorical variable. Naturally, this format imposes a logical constraint that enforces students from the same \texttt{School\_ID} must have identical values of school-level variables, such as \texttt{SchoolRegion}, \texttt{SchoolLocation}, and \texttt{SchoolClimate}. Preparing multilevel data in an appropriate format and accurately defining its schema are important tasks in the first stage because these decisions influence the choice of compatible GenAI models. 

\subsection{AI Training for Multilevel Data}\label{sec:AI_multilevel}

We train GenAI models using multilevel data in either a multiple-table format or a join-as-one format. Specifically, we train ClavaDDPM with multiple-table data, and we train CTGAN with join-as-one data. Below, we detail each GenAI model and enhance them with our targeted modifications for multilevel data generation.

\subsubsection{ClavaDDPM with Variable Decomposition}\label{sec:clava_vd}

ClavaDDPM \parencite{pang2024clavaddpm} is designed for synthesizing multi-relational (multi-table) datasets, ranging from two tables (e.g., student-school) to  three or more tables (e.g., student-school-district). In particular, ClavaDDPM utilizes classifier-guided diffusion models, which use an external classifier to steer the sampling process toward specific classes and address scalability challenges and long-range dependencies across multiple tables. For clarity, we use the term ``student table'' to refer to the child table, and ``school table'' to refer to the parent table.

The fundamental innovation of ClavaDDPM is the use of latent group labels (i.e., latent groups that combine original clusters) as intermediaries to model relationships between tables under foreign-key constraints (e.g., \texttt{School\_ID} in a student table). It introduces a latent group variable $g$ rather than directly modeling the conditional distribution $p(a_j \mid b_j)$, where $a_j$ represents student data belonging to school $j$ ($j=1,2,\dots,J$), and $b_j$ represents school $j$'s data. This approach is motivated by the fact that learning the distribution $p(a_j, b_j)$ is non-trivial due to the varying numbers of students in each school and a large number of schools. Thus, the model assumes that $a_j$ is independent of $b_j$ conditional on $g$: $a_j\indep b_j \mid g$. Using the low-dimensional latent variable $g$, the problem can be reformulated as \parencite{pang2024clavaddpm}:
\begin{align}
    p(a_j, b_j) = \sum_g p(a_j, b_j \mid g) p(g) = \sum_g p(a_j \mid  g) p(b_j \mid g) p (g) = \sum_g p (a_j \mid  g) p (b_j, g) \ .
\end{align}
Each $a_j$ models its size explicitly with a variable $n_j$, where $n_j$ is assumed to be conditionally independent of the row variable $a_{ij}$ for student $i$ given the latent variable $g$. Under this assumption, a generative process for $a_j$ is written as:
\begin{align*}
p(a_j \mid g) = p (n_j \mid g) \prod_{i=1}^{n_j} p(a_{ij} \mid g) \ .
\end{align*}
\indent However, although using the latent variable $g$ reduces the complexity in the generative process, it does not accurately recover the ICC between the student-school tables. As will be elaborated in our demonstration sections, the ICC often approaches zero under the current ClavaDDPM's architecture. In the social sciences, the ICC is one of the most important measures to summarize the clustering effect, and it is frequently used as a design factor in Monte Carlo simulation studies. To improve the model's ability to capture the ICC value, we propose a variable decomposition strategy that decomposes the $p$-th student-level variable $a_{ijp}$ into a school-demeaned variable $\widetilde{a}_{ijp}$ and a school-mean variable $\overline{a}_{jp}$ as: 
\begin{align*}
a_{ijp} &= ( a_{ijp} - \frac{\sum_{i=1}^{n_j} a_{ijp}}{n_j}) +  \frac{\sum_{i=1}^{n_j} a_{ijp}}{n_j} \\
       &= \widetilde{a}_{ijp} + \overline{a}_{jp} \ .
\end{align*}
Using such school-demeaned variables allows student-level variables to capture deviations from the school means within their respective schools, while their explicit school means are incorporated into the school table. Feeding this transformed input data into ClavaDDPM enables the model to explicitly learn within- and between-school variances and thus, preserve the original clustering effects without modifying the entire model architecture. This is a streamlined and computationally efficient solution to the ICC recovery problem. 

For convenience, we denote the school-demeaned student $i$'s row variable as $\widetilde{a}_{ij}$ and denote the corresponding school-mean row variable $\overline{a}_{j}$. We also denote student data from school $j$ after de-meaning as $\tilde{a}_j$, and denote the combination of existing school row $b_j$ and extra school-mean variables $\bar{a}_j$ as $\bar{b}_j=\{b_j, \bar{a}_j\}$. Importantly, we use $\tilde{a}_j$ and $\bar{b}_j$ instead of $a_j$ and $b_j$ as our input data for ClavaDDPM. Our modified generation process for student table $A$ and school table $B$ is formally written as:
\begin{align}
p(A, B) &\approx \prod_{j=1}^{J} p(\tilde{a}_j, \bar{b}_j) \nonumber \\
 & \approx \prod_{j=1}^{J} \sum_g p(n_j \mid g) \prod_{i=1}^{n_j} p(\tilde{a}_{ij}\mid g)p(\bar{b}_j, g) = \prod_{j=1}^{J} \sum_g p(\bar{b}_j, g)p(n_j \mid g) \prod_{i=1}^{n_j} p(\tilde{a}_{ij}\mid g) \ . \label{eq:ClavaDDPM_mod}
\end{align}
Specifically, in the first phase, we decompose student-level variables into school-demeaned and school-mean variables, including the former in the student table $A$ and the latter in the school table $B$. Then, we learn the latent variable $g$ on the joint space across tables, such that each school row $\bar{b}_j$ corresponds to a learned latent variable $g_j$. We augment the school $j$'s data into $(\bar{b}_j, g_j)$. In the second phase, we train the school diffusion model $p_\theta(\bar{b}, g)$ and the student diffusion model $p_{\phi}(\tilde{a})$. Given the learned latent $g$ and student table, we train classifier $p_{\psi}(g\mid \tilde{a})$. We then estimate the school size distributions conditioned on the latent variable $p(n_j\mid g)$. In the third phase, we synthesize the augmented school table from the trained school diffusion model. For each synthesized latent variable, we sample the school size, and given each synthesized school size, we sample each student row within each synthetic school. 

\subsubsection{CTGAN with Cluster-Level Consistency Alignments}\label{sec:CTGAN_cc}

CTGAN is a specialized GAN for synthesizing high-quality tabular data \parencite{xu2019modeling}. It introduces a conditional generation architecture for tabular data, effectively handling mixed data types (e.g., numerical, categorical), non-Gaussian and multi-modal distributions, and imbalance in categorical variables. CTGAN follows the standard adversarial framework \parencite{goodfellow2014generative} that consists of a generator ($G$) and a critic (or discriminator $D$). The critic aims to differentiate between real and synthetic data, whereas the generator strives to produce synthetic data that the critic cannot distinguish from real data. For training stability, CTGAN utilizes Wasserstein GAN with gradient penalty \parencite[WGAN-GP;][]{gulrajani2017improved}. The training process optimizes the minimax objective defined as:
\begin{align}
\min\limits_{G} \  \max\limits_{D \in \mathcal{D}} \  E_{x \sim \mathds{P}_r} [D(x)] - E_{\hat{x} \sim \mathds{P}_g} [D(\hat{x})] + \lambda E_{\tilde{x} \sim \mathds{P}_{\tilde{x}}}[\{\|\nabla_{\tilde{x}} D(\tilde{x})\|_2 - 1\}^2].
\label{eq:minmax}
\end{align}
\noindent Here, $\mathcal{D}$ represents the family of 1-Lipschitz functions. The term $\mathds{P}_r$ denotes the real data distribution, while $\mathds{P}_g$ denotes the generator's distribution defined by $\hat{x}=G(z)$, where $z$ is sampled from a noise distribution (e.g., a Gaussian distribution). The term $\tilde{x}$ denotes a linear interpolation between real and synthetic data, $\tilde{x}=\rho x + (1-\rho)\hat{x}$, with $\rho$ drawn from a uniform distribution. The term $\lambda$ represents the weight for the gradient penalty (typically 10). In this framework, the critic $D$ seeks to maximize the Wasserstein distance between real and synthetic distributions, subject to a gradient penalty that enforces the Lipschitz constraint, whereas the generator $G$ minimizes this distance. The two networks are iteratively updated in an adversarial manner until convergence \parencite{goodfellow2014generative, xu2019modeling}. See \Textcite{xu2019modeling} for further details on key techniques for generating realistic tabular data, including mode-specific normalization, one-hot encoding, and training-by-sampling.

Since CTGAN is inherently designed for single-table generation, it does not automatically preserve the hierarchical structures of multilevel data. Without explicit constraints, CTGAN may generate logical inconsistencies where students within the same school are assigned different values for school-level variables. To rectify this, we could use a simple ad-hoc constraint that enforces the same values for school-level variables within schools. That is, if synthetic students in the same school show varying school-level values, we replace these values with a measure of central tendency (e.g., mode, mean, or median), depending on the variable's measurement scale. 

However, strictly forcing values toward the central tendency often distorts the marginal distributions of the synthetic data, and we frequently observe this problem for both categorical and numerical variables. Although the raw synthetic data may initially match the real data, enforcing within-cluster consistency through naive aggregation can substantially compromise fidelity. To address this issue, we introduce a new post-processing procedure, \textit{cluster-level consistency alignment} (CLCA). 
CLCA enforces identical school-level values within each synthetic school while preserving fidelity to the real data through separate alignment procedures for numerical and categorical variables. We choose specific alignment procedures based on computational efficiency and algorithmic simplicity. For numerical school-level variables, CLCA aligns synthetic school-level summaries to the empirical school-level distribution in the real data through a rank-based quantile matching procedure, which corresponds to one-dimensional optimal transport. For categorical school-level variables, CLCA solves a constrained optimization problem for the school-level assignment. Let $x_{j,k}\in\{0,1\}$ indicate whether school $j$ is assigned to category $k$ ($k=1, \ldots, K$), with the constraint $\sum_{k=1}^{K} x_{j,k}=1$ for all $j$. The assignment is chosen to minimize the following objective:
\begin{align}
\alpha \sum_{k=1}^{K} \left| \frac{1}{N}\sum_{j} n_j x_{j,k} - p_k^* \right|
+ \beta \sum_{k=1}^{K} \left| \sum_{j} x_{j,k} - t_k^* \right|
+ \gamma \sum_{j}\sum_{k=1}^{K} \bigl(-\log \hat{p}_{j,k}\bigr)x_{j,k} \ .
\end{align}
Here, $p_k^*$ denotes the student-level proportion for category $k$ in the real data, and $t_k^*$ denotes the number of schools in category $k$ in the real data. For the synthetic data, $\hat{p}_{j,k}$ denotes school-specific proportion for category $k$, $n_j$ is the size of school $j$, and $N=\sum_j n_j$. The parameters $\alpha$, $\beta$, and $\gamma$ represent weights for each term, and in our implementation, they are all set to 1. This objective balances fidelity to the real student-level marginals with fidelity to the real school-level category counts, while enforcing consistency of school-level variables within each school. We solve the optimization problem using the CP-SAT solver in Google OR-Tools \parencite{cpsatlp}. See Appendix~\ref{app:cc} for detailed CLCA algorithms for numerical and categorical variables.

\subsection{Quality Evaluation for Synthetic Multilevel Data}\label{sec:qual_eval}

In this section, we summarize formal metrics and visualization techniques to check within-table fidelity (how closely synthetic data resemble real data within each table), between-table fidelity (how closely synthetic data resemble real data between tables), or ML efficacy (how similarly ML models perform on synthetic data compared to real data). We leverage existing formal metrics and introduce new metrics based on the statistics and multilevel modeling literature. These formal metrics typically range from 0 to 1, where 0 indicates no similarity and 1 indicates perfect similarity. While there is no universal consensus on fixed thresholds for synthetic data similarity scores, both open-source benchmarks (e.g., SDV) and industry standards (e.g., Gretel.ai, Mostly AI) provide heuristic guidance. Table \ref{tab:scale_inter} summarizes heuristic cutoffs for interpreting similarity scores based on the existing literature \parencite[e.g.,][]{zhang2022sequential} and our own experiences. 

\begin{table}[!ht]
\centering
  \caption{Heuristic interpretation for a generic 0-1 similarity score}
  \label{tab:scale_inter}
  \small
    \begin{threeparttable}
    \begin{tabular}{l L{2.6cm} L{10cm}}
\toprule
    Score range & Category & Interpretation \\
\midrule
    0.90 -- 0.99 &  Excellent \newline (High) & Synthetic data are statistically very similar to real data. \\

    0.80 -- 0.89 & Good \newline (Medium-high) &  Synthetic data capture major distributions of real data.\\

    0.60 -- 0.79 & Fair \newline (Medium-low) &  Basic variable distributions are similar, but complex and local patterns may not be preserved. \\

    < 0.60 & Poor \newline  (Low) & Synthetic data differs significantly from real data. \\
\bottomrule         
    \end{tabular}
\vspace{1mm}
  \begin{tablenotes}[para,flushleft]
    \footnotesize NOTE: Scores close to 1.0 (or $\geq$ 0.99) may indicate overfitting where the generator memorizes real data.
  \end{tablenotes}
  \end{threeparttable}   
\end{table}

\subsubsection{Within-Table Fidelity}

We measure \textit{marginal} and \textit{pairwise} fidelity within each table. Marginal fidelity metrics evaluate whether the distributions of individual variables in synthetic data are comparable to those of their real counterparts, in terms of mean, variance, counts, or distribution shape. Pairwise fidelity metrics capture the similarity of relationships between variable pairs in synthetic data versus real data. Table \ref{tab:within_fidel} summarizes the visualization tools and formal metrics to inspect within-table fidelity. We denote a marginal substructure measure (e.g., cumulative distribution functions, probability mass functions) of the synthetic data distribution $\mathds{P}_s$ and real data distribution $\mathds{P}_r$ as $p^{m}_s$ and $p^{m}_r$, respectively. In general, a metric is formulated as $1-\text{dist}(p^{m}_s, p^{m}_r)/c$, where $\text{dist}(\cdot)$ denotes a distance function that quantifies the difference between the synthetic and real distributions, and $c$ is a normalizing factor so that the metric ranges from 0 to 1.

\begin{table}[!ht]
\centering
  \caption{Metrics for Within-Table Fidelity}
  \label{tab:within_fidel}
  \small
    \begin{threeparttable}
    \begin{tabular}{L{2.5cm} L{4cm} L{4.5cm} l }
\toprule
  Data Type & Visualization & Metric     & Equation   \\
\midrule
    \multicolumn{2}{l}{\textit{Marginal fidelity}} & \\  

  Numeric  & Histogram, density, boxplot, etc. & KSC & $1-d_{\rm KS}(p^{m}_s,  p^{m}_r)$   \\
  Categorical & barplot, piechart, etc. &  TVC & $1-d_{\rm TV}(p^{m}_s,  p^{m}_r)$  \\
    \hline
    \multicolumn{2}{l}{\textit{Pairwise fidelity}} & \\    
  Num-Num &  Scatterplot, etc. & Correlation similarity  & $1-|p^{b}_s -   p^{b}_r|/2$  \\  
  Cat-Cat & Mosaicplot, etc. &  Contingency similarity   & $1-d_{\rm TV}(p^{b}_s,  p^{b}_r)$\\  
  Num-Cat   &  Grouped  histogram,   &    Contingency similarity  & $1-d_{\rm TV}(p^{b}_s,  p^{b}_r)$    \\
 & density, boxplot, etc.  & Eta-squared similarity$^{\star}$  & $1-|\eta^{2}_s -   \eta^{2}_r|$  \\
 Both-Both   & & MI similarity  &  $1-d_{\rm TV}(M_s, M_r)$   \\ 
\bottomrule         
    \end{tabular}
\vspace{1mm}
  \begin{tablenotes}[para,flushleft]
    \footnotesize NOTE: KSC = Kolmogorov-Smirnov complement; TVC = Total Variation  Distance complement; MI = Mutual Information; $^{\star}$ = newly introduced metric in this paper.
  \end{tablenotes}
  \end{threeparttable}   
\end{table}

For numeric variables, we use histograms, density plots, box plots, or violin plots to visually compare their marginal distributions. We also use a formal metric known as the \textit{Kolmogorov-Smirnov complement} (KSC), which is defined as $1 - d_{\rm KS}(p^{m}_s, p^{m}_r)$, where $d_{\rm KS}$ represents the Kolmogorov-Smirnov statistic. This statistic measures the maximum difference between the empirical cumulative distribution functions of synthetic and real variables. For categorical variables, we use bar plots or pie charts for visual inspection and use a metric known as the \textit{Total Variation  Distance complement} (TVC). TVC is defined as $1-d_{TV}(p^{m}_s, p^{m}_r)$, where $d_{\rm TV}$ represents the total variation distance, which is half the sum of the absolute differences in proportions between the categories of the synthetic variable and the real variable \parencite{sdmetrics, suk2025using}.

Next, we evaluate pairwise/bivariate relationships. We denote a bivariate substructure measure (e.g., correlation coefficient, contingency table proportions) of the synthetic and real data distributions as $p^{b}_s$ and $p^{b}_r$, respectively. For numeric variable pairs, we use scatter plots and a \textit{correlation similarity} metric , which is written as $1-|p^{b}_s - p^{b}_r|/2$. This metric calculates the complement of half the absolute difference between the correlation coefficients of the synthetic data and real data. Similarly, for relationships between two categorical variables, we use mosaic plots and a \textit{contingency similarity} metric that measures the total variation distance between the contingency table proportions of the synthetic data and real data \parencite{sdmetrics, suk2025using}. 

Furthermore, for relationships between a numeric variable and a categorical variable, we can draw per-category histograms, density plots, box plots or apply the contingency similarity metric after binning the numeric variable. However, it is well known that the efficacy of binning largely depends on the number and location of bins. To better reflect the specific data types, we propose using an effect size measure from ANOVA called Eta-squared ($\eta^2$). The Eta-squared is defined as $\eta^2=SS_{\rm between}/SS_{\rm total}$, where $SS_{\rm between}$ is the sum of squares between categories (i.e., the variation in the numeric variable explained by the categorical variable) and $SS_{\rm total}$ is the total sum of squares \parencite{cohen1988statistical}. The \textit{Eta-squared similarity} is defined as:
\begin{align}\label{eq:eta}
\text{Eta-squared similarity} = 1-|\eta^{2}_s-\eta^{2}_r|. 
\end{align}
This metric ranges from 0 to 1 and naturally captures the strength of the relationship between a numeric and a categorical variable with two or more categories.

Additionally, \textit{Mutual Information (MI) similarity} serves as a global measure to quantify both linear and non-linear dependencies between any pair of variables \parencite{yang2024structured}. MI measures the reduction in uncertainty about one variable given another variable. The MI similarity score is constructed as follows: first, the raw MI is computed for every pair of columns in both synthetic and real data; second, the diagonal elements are set to 0, and the entire matrix is normalized so that the sum of all elements equals 1. Then the MI similarity is computed as \parencite{yang2024structured}:
\begin{align}\label{eq:MI}
\text{MI similarity} = 1 - d_{\text{TV}}(M_s, M_r) \ .
\end{align}
Here, $M_s$ and $M_r$ represent the normalized MI matrices for synthetic data and real data, respectively. The total variation distance is computed across all cell pairs for the two matrices. 

Lastly, the average of either the marginal fidelity scores or the pairwise fidelity scores is computed to summarize similarity within each subdomain. Likewise, the average of all within-table fidelity scores can be reported to provide an overall summary of similarity for each table.

\subsubsection{Between-Table Fidelity}

Beyond within-table fidelity, we leverage existing metrics and establish new metrics to assess the quality of relationships between multiple tables (e.g., between student-level and school-level tables). An existing metric, \textit{referential integrity}, examines the extent to which foreign key (i.e., parent/school ID) value in a student table exists as a primary key value in the corresponding school table (e.g., every \texttt{school\_ID} value in the student table exists in the school table). Another existing metric, \textit{cardinality shape similarity}, focuses on the distribution of cluster size $n_j$ (e.g., the number of students per school). It assesses whether the cluster size distribution in synthetic data are statistically similar to that of real data, using the KSC metric \parencite{sdmetrics, pang2024clavaddpm}. Additionally, the \textit{$k$-hop correlation} metric employs the aforementioned correlation similarity metric to quantify the relationship between two numeric variables across tables: one from a school table and the other from a student table. Here, $k$ denotes the relational distance between the target tables $(k=1,2,3, \ldots)$.\footnote{Note that $k=0$ would indicate two variables within the same table, which falls under within-table fidelity.} To our knowledge, the prior literature has focused on $k$-hop analysis using only  numeric variables \parencite[e.g.,][]{pang2024clavaddpm}. See Table \ref{tab:btw_fidel} for a summary of between-table fidelity metrics.

\begin{table}[!ht]
\centering
  \caption{Metrics for Between-Table Fidelity}
  \label{tab:btw_fidel}
  \small
    \begin{threeparttable}
    \begin{tabular}{l L{10.5cm}}
\toprule
    Metric & Description \\
\midrule
    Referential integrity &  Proportion of parent IDs matched across tables \\
    Cardinality shape similarity & Similarity of the number of individuals per cluster  \\
    $k$-hop correlation similarity & Similarity of the correlations of numeric variables from tables at level distance $k$  \\
    $k$-hop contingency similarity$^{\star}$ & Similarity of the dependencies of categorical variables from tables at level distance $k$   \\
    $k$-hop Eta-squared similarity$^{\star}$ & Similarity of the dependencies between a numeric variable and a categorical variable from tables at level distance $k$  \\     
    $k$-hop MI similarity$^{\star}$ & Similarity of the dependencies of any-type variables from tables at level distance $k$  \\            
    ICC  similarity$^{\star}$  & Similarity of intra class correlation (ICC) \\ 
    Reliability  similarity$^{\star}$  &  Similarity of the average cluster-mean reliability  \\  
\bottomrule         
    \end{tabular}
\vspace{1mm}
  \begin{tablenotes}[para,flushleft]
    \footnotesize NOTE: $k$-hop = $k$-level distance between the target parent and child tables ($k=1,2,\ldots$); MI = Mutual Information;  $^{\star}$ = newly introduced metric in this paper. 
  \end{tablenotes}
  \end{threeparttable}   
\end{table}

To fully capture long-range dependencies, we extend the pairwise fidelity metrics listed in Table \ref{tab:within_fidel} for $k$-hop analysis. We introduce \textit{$k$-hop contingency similarity} to investigate the similarity of contingency table proportions for two categorical variables that come from different structural levels. Likewise, \textit{$k$-hop Eta-squared similarity} measures the dependency between a numeric variable and a categorical variable from different levels, and \textit{$k$-hop MI similarity} captures the dependencies between any pair of variables across levels. These $k$-hop metrics evaluate long-range dependencies across the data hierarchy. 

Moreover, we develop new metrics to capture between-table fidelity by drawing insights from the multilevel modeling literature \parencite[e.g.,][]{raudenbush2002hierarchical, snijders2012multilevel}. One of the most well-known measures for summarizing the relationship between school and student tables is ICC. The ICC is defined as the proportion of total variance explained by between-cluster variation and measures the similarity of individuals' outcomes within the same cluster compared to those in other clusters. Formally, the ICC is computed as:
\begin{align}
 \text{ICC} &= \frac{\tau^2}{\tau^2+\sigma^2} \label{eq:ICC} \ , 
\end{align}
where $\tau^2$ and $\sigma^2$ represent between-cluster variance (school level) and within-cluster variance (student level) of each target variable, respectively. For binary variables,  $\sigma^2$ in Equation \eqref{eq:ICC} is set to $\frac{\pi^2}{3}$, the assumed variance of the standard logistic distribution. The ICC ranges from no correlation ($\text{ICC}$ = 0) to complete correlation ($\text{ICC}$ = 1). For multicategorical variables, the ICC can be computed for each category's dummy variable, and then the category-specific ICC values are aggregated \parencite{nakagawa2010repeatability}. 

In the social sciences, the ICC values observed in real data are typically not high, often ranging from 0.05 to 0.3 \parencite{aguinis2013best}. In this case, a simple absolute difference between real and synthetic ICC values can yield misleadingly small numbers that do not meaningfully differentiate data quality. Therefore, we propose the \textit{ICC similarity} metric: 
\begin{align}\label{eq:ICCsim}
\text{ICC similarity} = \left(1-|\text{ICC}_s - \text{ICC}_r|\right)^{3}.
\end{align}
Here, $\text{ICC}_s$ and $\text{ICC}_r$ represent the ICC values obtained from synthetic data and real data, respectively. The cubic transformation is applied to amplify small absolute differences so that the metric better differentiates the varying quality of synthetic data.\footnote{We explored alternative ICC similarity formulas, but this version consistently captured quality differences between synthetic datasets produced by different generators.} We apply the ICC similarity metric \eqref{eq:ICCsim} to each target variable in the dataset. 

Lastly, we propose a formal metric based on the reliability of the cluster means of an outcome, which is another popular measure in the multilevel modeling literature. The average cluster-mean reliability, denoted as Rel, and the corresponding \textit{reliability similarity} are written as:
\begin{align}
 \text{Rel} &= \frac{\bar{n}\tau^2}{\bar{n}\tau^2+\sigma^2}, \\
 \text{Reliability similarity} &= 1 - |\text{Rel}_s - \text{Rel}_r|.
\end{align}
Here, $\bar{n}$ represents the average cluster size, and $\text{Rel}_s$ and $\text{Rel}_r$ represent the average cluster-mean reliability in synthetic data and real data, respectively. Similar to the ICC, we replace $\sigma^2$ with $\frac{\pi^2}{3}$ for binary variables. As with the above metrics, this metric ranges from 0 to 1, where higher values indicate greater similarity. The average of these between-table fidelity measures can be reported to summarize the overall structural quality of the synthetic data.

\subsubsection{ML Efficacy}

In addition to data fidelity, we assess ML efficacy as a quality metric by comparing the utility of ML models trained on synthetic and real data. In each iteration, we select one variable as the target and train ML models (e.g., random forests, XGBoost) using the remaining variables as predictors. This training is performed separately on the synthetic and real training datasets. We then evaluate the performance of these models on the same held-out dataset of real data. This procedure follows the widely used ``Train on the Synthetic and Test on the Real'' (TSTR) and ``Train on the Real and Test on the Real'' (TRTR) evaluation \parencite{hernandez2025comprehensive}. To stabilize the results, we apply five-fold cross-validation. 

Furthermore, to account for the multilevel data structure in ML efficacy, we employ ML models tailored for multilevel data, such as linear mixed-effects regression \parencite{Bates2015Fitting} and Gaussian Process Boosting \parencite{Fabio2022Gaussian}. For numeric target variables, we perform regression and evaluate the $R^2$ score, and for categorical target variables, we perform classification and evaluate the $F1$ score (or other classification metrics). We measure ML efficacy by computing the absolute difference in the chosen performance metric, $m$, between the TSTR and TRTR processes. Formally, 
\begin{align}\label{eq:mle}
\text{ML Efficacy} = 1 - |m_{\rm TSTR} - m_{\rm TRTR}|,
\end{align}
where $m$ represents the $R^2$ score for numerical variables and $F1$ score for categorical variables. The overall ML efficacy can be evaluated by averaging variable-specific efficacy scores.

We remark that there are other dimensions of quality evaluation, such as \textit{generalization}---whether the generated sample is not a mere copy of existing data \parencite{alaa2022faithful, suk2025using}. We can check generalization by measuring the proportion of whether each row in the synthetic data is new or matches an existing row in the real data. Quality synthetic data should achieve one in the generalization score, indicating that all synthetic rows are new. See \textcite{suk2025using} for more details on generalization.

\section{Demonstration 1: Predictive Performance}\label{sec:demo1}

In this section, we demonstrate our AI-based simulation approach for evaluating the predictive performance of statistical methods, and we consider the prediction rules studied by \Textcite{afshartous2005prediction} as an illustrative example. 

\subsection{Stage 1: Specifying quantitative methods of interest and data schema} 

We consider the three prediction rules investigated by \Textcite{afshartous2005prediction} for two-level data, where individuals (e.g., students) are nested within clusters (e.g., schools). These rules are (i) ordinary least squares with cluster-specific fixed intercepts and slopes, denoted as OLS$^{\rm FE}$, (ii) the prior method, which uses only the fixed-effects components from a random-effects model, and (iii) the multilevel method, which uses both fixed- and random-effects components from a random-effects model. 

Since \Textcite{afshartous2005prediction} did not include a real data application, we utilize a nationally representative educational dataset extensively used in educational research: the High School Longitudinal Study of 2009 \parencite[HSLS,][]{ingels2011high}. Following \Textcite{pan_learning_2026}, we focus on 17 student-level variables (e.g., student ID, gender, race/ethnicity, math identity) and 11 school-level variables (e.g., school ID, school type, school climate). The outcome of interest is the student's standardized math achievement score. After standard data cleaning procedures, our analytical sample comprises 17,042 students from 823 schools. Based on the backward stepwise selection algorithm, we select 12 variables that are directly relevant to the prediction task. Additionally, we prepare both multiple-table and join-as-one versions of the dataset, as illustrated in Figure~\ref{fig:muldata}. See Appendix~\ref{app:HSLSdata} for additional details on the input data schema.

\subsection{Stage 2: Training GenAI models based on real data}\label{sec:demo1_step2}

We train ClavaDDPM on the HSLS dataset in a multiple-table format. As discussed in Section \ref{sec:clava_vd}, we apply the variable decomposition strategy where student-level variables are decomposed into cluster-demeaned and cluster-constant components, and these components replace the original variables in the dataset. For hyperparameter tuning, we set the number of latent groups to $k=50$ while maintaining the default values for the other parameters (e.g., 2000 diffusion timesteps and a learning rate of 0.0006).

We also train CTGAN on the HSLS dataset in a join-as-one format. The generator network uses two hidden layers with 256 units each and rectified linear unit (ReLU) activation functions. The discriminator network uses two hidden layers with 256 units each, but with a modified version of ReLU known as Leaky ReLU. The learning rates are both set to 0.0002, and the model is trained for 1600 epochs. We keep the remaining CTGAN settings at their default values (e.g., a batch size of 500 and an embedding dimension of 128).

\subsection{Stage 3: Checking the quality of synthetic data}\label{sec:demo1_step3}

\begin{figure}[!b]
    \centering
    \includegraphics[width=\linewidth]{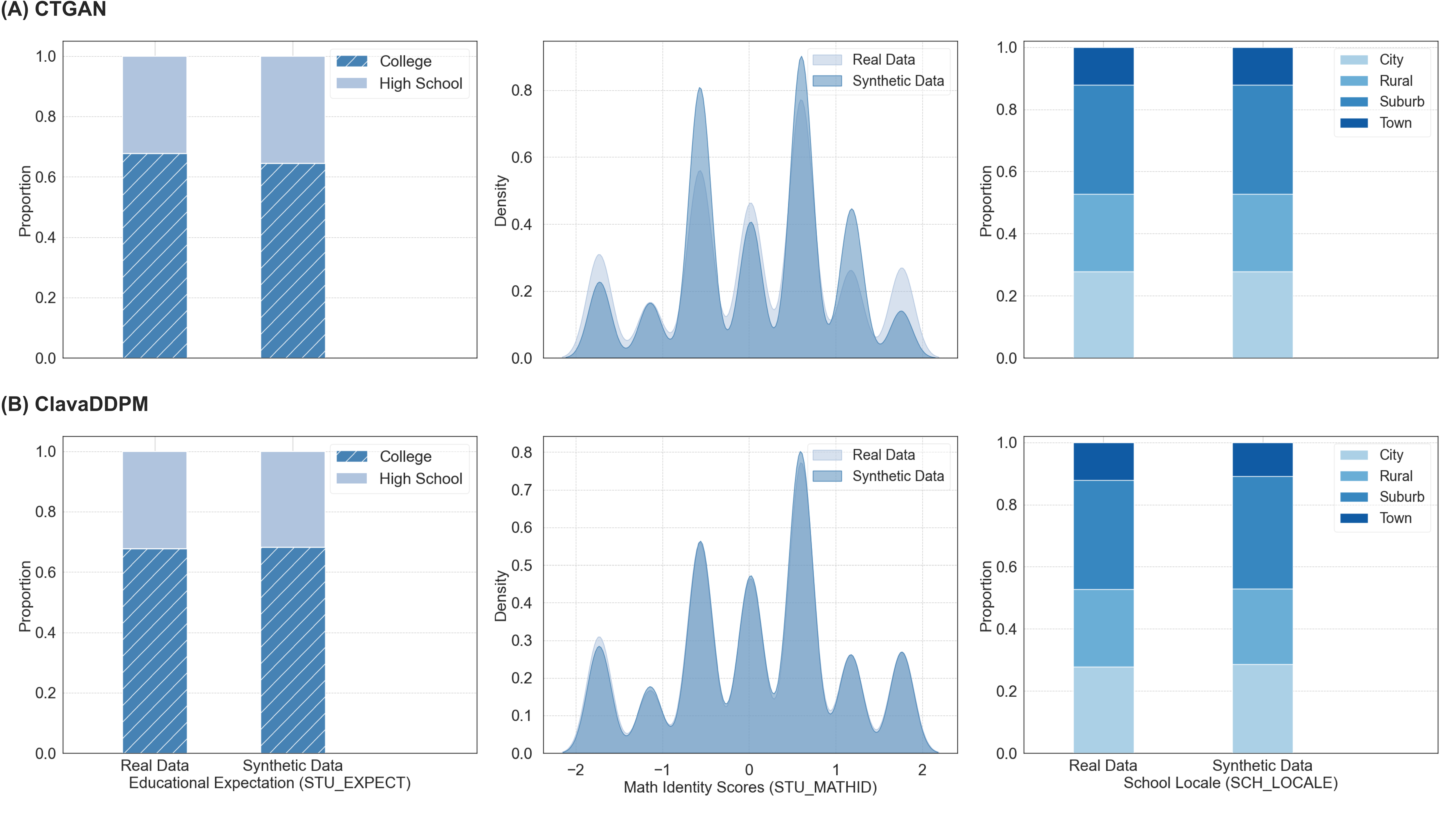}
    \caption{Visual checks on selected student- and school-level variables.}
    \label{fig:visual}
\end{figure}

For quality evaluation, we use the trained ClavaDDPM model to generate synthetic datasets that have the same number of schools as the original dataset ($J = 823$). In contrast, we use the trained CTGAN model to generate synthetic datasets with the same number of students as the original dataset ($N = 17,042$). We then apply CLCA to enforce consistency of school-level variables, as described in Section~\ref{sec:CTGAN_cc}. For robust evaluation, we generate 10 candidate synthetic datasets per model and assess their average quality rather than using a single instance.

Figure \ref{fig:visual} visualizes the marginal distributions of two student-level variables (educational expectation and math identity) and one school-level variable (school locale) from the real dataset and each AI-generated synthetic dataset. While both CTGAN and ClavaDDPM generally produce synthetic data that closely mimic real data, ClavaDDPM demonstrates superior performance in capturing the real distribution of the math identity score. 

\begin{figure}[!htb]
    \centering
    \includegraphics[width=\linewidth]{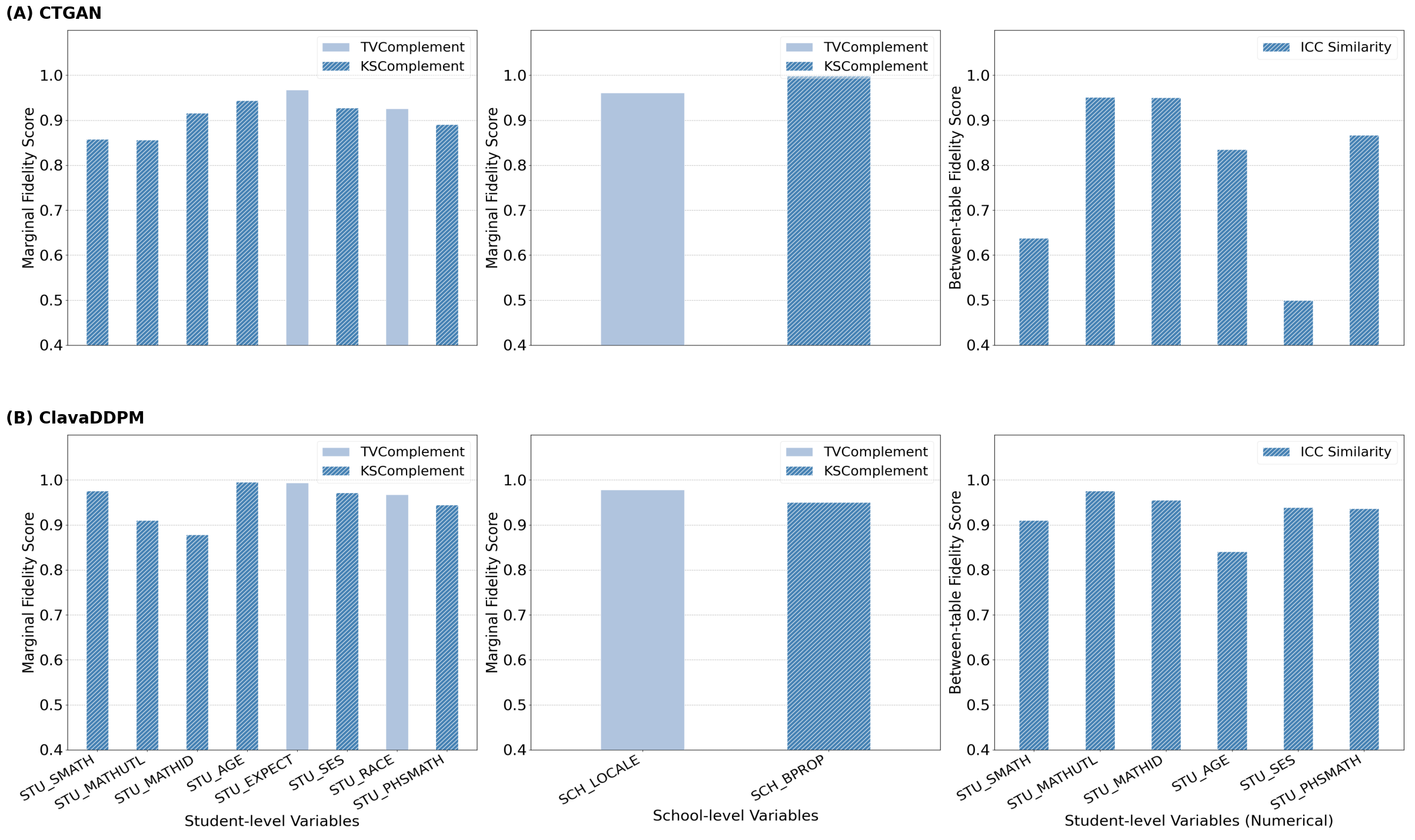}
    \caption{Quality results on selected formal metrics for CTGAN (top) and ClavaDDPM (bottom). TVC = Total Variation Distance complement; KSC = Kolmogorov-Smirnov complement; ICC = Intra-class correlation}
    \label{fig:quality_formal_demo1}
\end{figure}

Figure \ref{fig:quality_formal_demo1} summarizes the results for both within-table (assessed at both student- and school-level) and between-table fidelity. We focus on the within-table marginal fidelity metrics and our proposed between-table ICC similarity metric. For details of the other fidelity metrics and visual checks, see Appendix \ref{app:quality_evaluation}. 
From the figure, both GenAI models achieve high marginal fidelity for student- and school-level variables, with ClavaDDPM showing better student-table marginal fidelity. However, the main difference appears in between-table fidelity. ClavaDDPM shows excellent ICC similarity scores across student-level numeric variables, while CTGAN performs poorly on \texttt{STU\_SMATH} and \texttt{STU\_SES}. This pattern suggests that ClavaDDPM better preserves the hierarchical dependence structure in the original multilevel data. 

Regarding ML efficacy, CTGAN and ClavaDDPM achieve good to excellent performance across most student- and school-level variables. Specifically, ClavaDDPM achieves average ML efficacy scores of 0.963 and 0.825 for student- and school-level variables, respectively, whereas CTGAN yields a slightly lower score at the student level (0.937) and a slightly higher score at the school level (0.837). See Appendix~\ref{app:quality_evaluation} for more details.\footnote{Note that the generalization scores for both models are exactly 1, which means all observations in the synthetic data are new ones, rather than simply bootstrapping from the observed real data.}

Overall, ClavaDDPM faithfully preserves the clustering structure of multilevel data while capturing high fidelity in each table. Therefore, we use the ClavaDDPM-generated synthetic datasets only for the downstream simulation tasks.

\subsection{Stage 4: Designing a simulation procedure and conducting simulations}\label{sec:demo_step4}

This stage details the AI-based simulation procedure to evaluate the predictive performance of the three prediction rules investigated by \textcite{afshartous2005prediction}.

\begin{enumerate}
    \item \textit{Set simulation conditions}. Following \textcite{afshartous2005prediction}, we vary the number of schools $J$ across five conditions: 10, 25, 50, 100, and 300. 
    \item \textit{Select performance criteria}. We use predictive mean squared error (PMSE) to evaluate predictive accuracy, as in \textcite{afshartous2005prediction}.
    \item \textit{Select the number of replications}. The number of replications is set to 500 per condition.
    \item \textit{Generate synthetic datasets from GenAI models}. Under each simulation condition, we generate 500 synthetic datasets using the trained ClavaDDPM model, which inherently learns and preserves the ICC and cluster-size distributions from the real data. 
    \item \textit{Apply quantitative methods of interest to synthetic datasets}. We apply the three prediction rules (OLS$^{\rm FE}$, the prior method, the multilevel method)  to each synthetic dataset. Detailed model specifications are provided in Appendix~\ref{app:model_specification}.
    \item \textit{Measure predictive performance under the chosen criteria}. We measure the PMSE of the three methods across all synthetic datasets.   
\end{enumerate}

Additionally, we replicate the traditional simulation design used by  \textcite{afshartous2005prediction} to compare its results with those from our AI-based approach. Following their simulation design, we generate a simple data structure that consists of one student-level predictor, one school-level predictor, and a single outcome, all of which are drawn from normal distributions with no dependence between the predictors. Among the various simulation conditions in \textcite{afshartous2005prediction}, we select a baseline condition that most closely aligns with our empirical AI-based simulation design: an ICC of 0.2 and a cluster size of 25.

\subsection{Stage 5: Evaluating method performance} 

\begin{figure}[!ht]
    \centering
    \includegraphics[width=\linewidth]{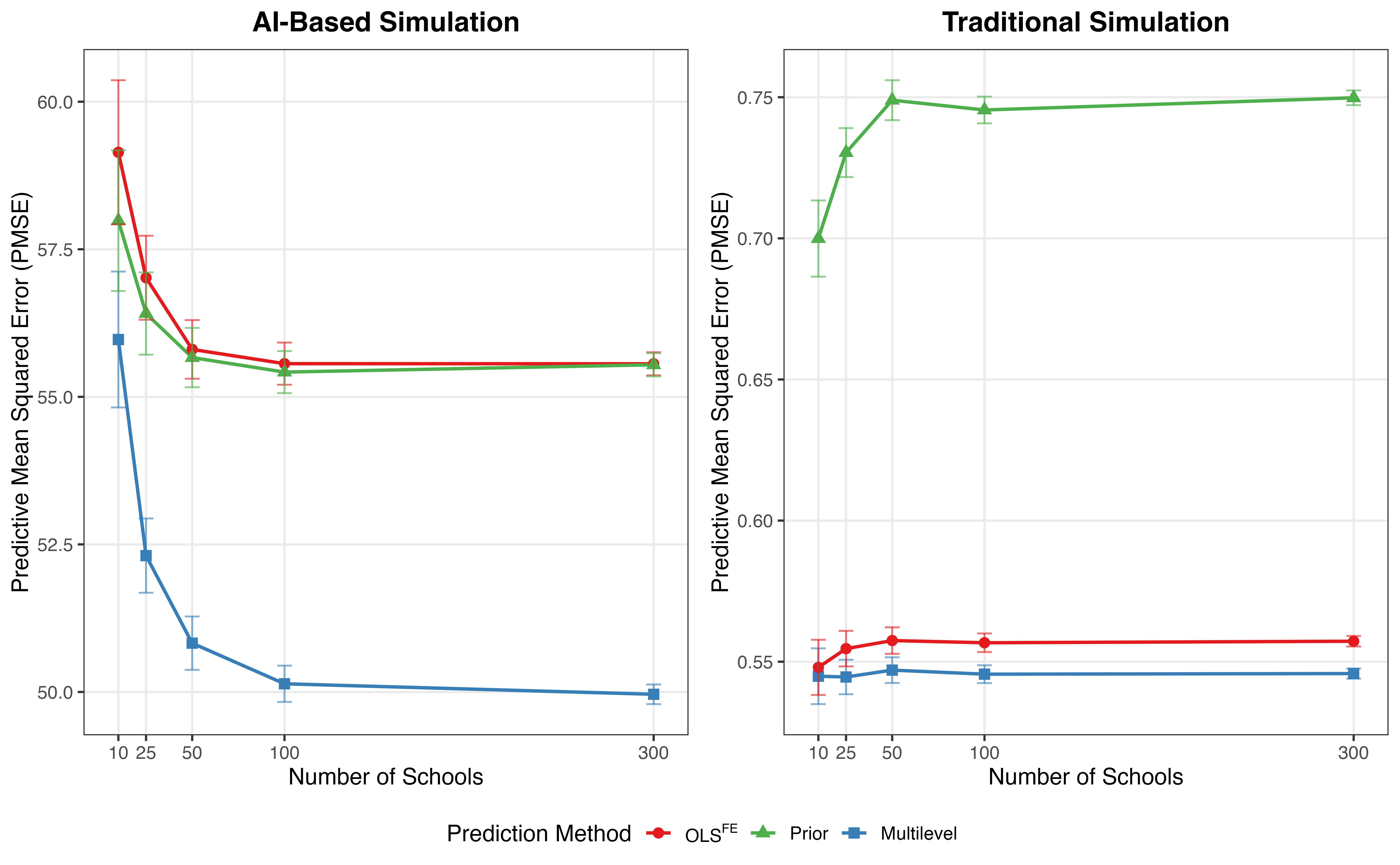}
    \caption{Results on the predicted mean squared error under the AI-based and traditional simulations.}
    \label{fig:Prediction}
\end{figure}

Figure \ref{fig:Prediction} summarizes the PMSE results of the three prediction rules from both AI-based and traditional simulations. In the AI-based simulation, the multilevel method achieves a significantly lower PMSE than the other two methods, and the PMSE values decrease as the number of schools $J$ increases across all three methods. In the traditional simulation, the multilevel method again produces the lowest PMSE. However, we observe a relatively constant PMSE for the multilevel and OLS rules, while the prior rule shows worsening performance as the number of schools increases. These results are consistent with the original findings of \textcite{afshartous2005prediction}. 

The divergent PMSE trends likely arise from differences in model specification and data complexity. In the traditional simulation with only one student-level predictor and one school-level predictor, OLS$^{\rm FE}$ includes school-specific intercepts and slopes of the student-level predictor, and the multilevel method incorporates random effects for both. As a result, these two methods show similar performance, whereas the prior method without using cluster-specific random effects performs the poorest. In contrast, in our AI-based simulation that uses seven student-level predictors and two school-level predictors, OLS$^{\rm FE}$ includes school-specific intercepts and, if applicable, school-specific slopes of only one of seven student-level predictors. Thus, the predictive gains from school-specific components in OLS$^{\rm FE}$ are limited in our realistic settings, resulting in accuracy levels similar to those of the prior method. Furthermore, the overall PMSE values in the AI-based simulation are much higher than those in the traditional simulation. This is expected as AI-generated synthetic data preserve real-world complexities among variables and incorporate a larger set of predictors, thus increasing the overall prediction error.

Finally, we remark that because CTGAN did not produce high-quality synthetic multilevel data (as demonstrated in Stage 3), we excluded it from our main AI-based simulations. However, for comparison purposes, we conducted an additional simulation using CTGAN-generated synthetic data. In this simulation, the three prediction rules behave almost identically. This highlights the importance of ensuring high data fidelity before designing and conducting an AI-based simulation study. See Appendix \ref{app:d1_CTGAN} for more details.

\section{Demonstration 2: Parameter Recovery}\label{sec:demo2}

In this section, we demonstrate our AI-based simulation approach for evaluating the parameter recovery of statistical methods, using the models investigated in \Textcite{huang2017multilevel} as an illustrative example.

\subsection{Stage 1: Specifying quantitative methods of interest and data schema}

For this demonstration, we examine four models of interest in \Textcite{huang2017multilevel}: (i) ordinary least squares (OLS) without cluster dummy variables, (ii) OLS including group means of student-level variables (OLS+IGM), (iii) hierarchical linear models (HLM; also known as random-effects or multilevel models), and (iv) HLM including group means of student-level variables (HLM+IGM). 

For our empirical data, we continue to use the HSLS dataset with 17,042 students from 823 schools. We select six student-level variables and three school-level variables that exhibit cross-level correlations. Notably, one school-level variable, \texttt{SCH\_CLIMATE}, is significantly correlated with both a student-level variable \texttt{STU\_SES} ($\rho=0.19$) and another school-level variable \texttt{SCH\_ASSIST} ($\rho=-0.21$). The outcome variable is students' ninth-grade academic GPA. See Appendix~\ref{app:HSLSdata} for model details.

\subsection{Stage 2: Training GenAI models based on real data}  

Since ClavaDDPM demonstrated superior performance over CTGAN in terms of both data fidelity and ML efficacy, and the between-table fidelity results of CTGAN were unsatisfactory in Section~\ref{sec:demo1}, we exclude CTGAN in this demonstration. We train ClavaDDPM on the HSLS data in a multiple-table format using the variable decomposition strategy, following the same procedure described in Section~\ref{sec:demo1_step2}.

\subsection{Stage 3: Checking the quality of synthetic data}  

\begin{figure}[!ht]
    \centering
    \includegraphics[width=\linewidth]{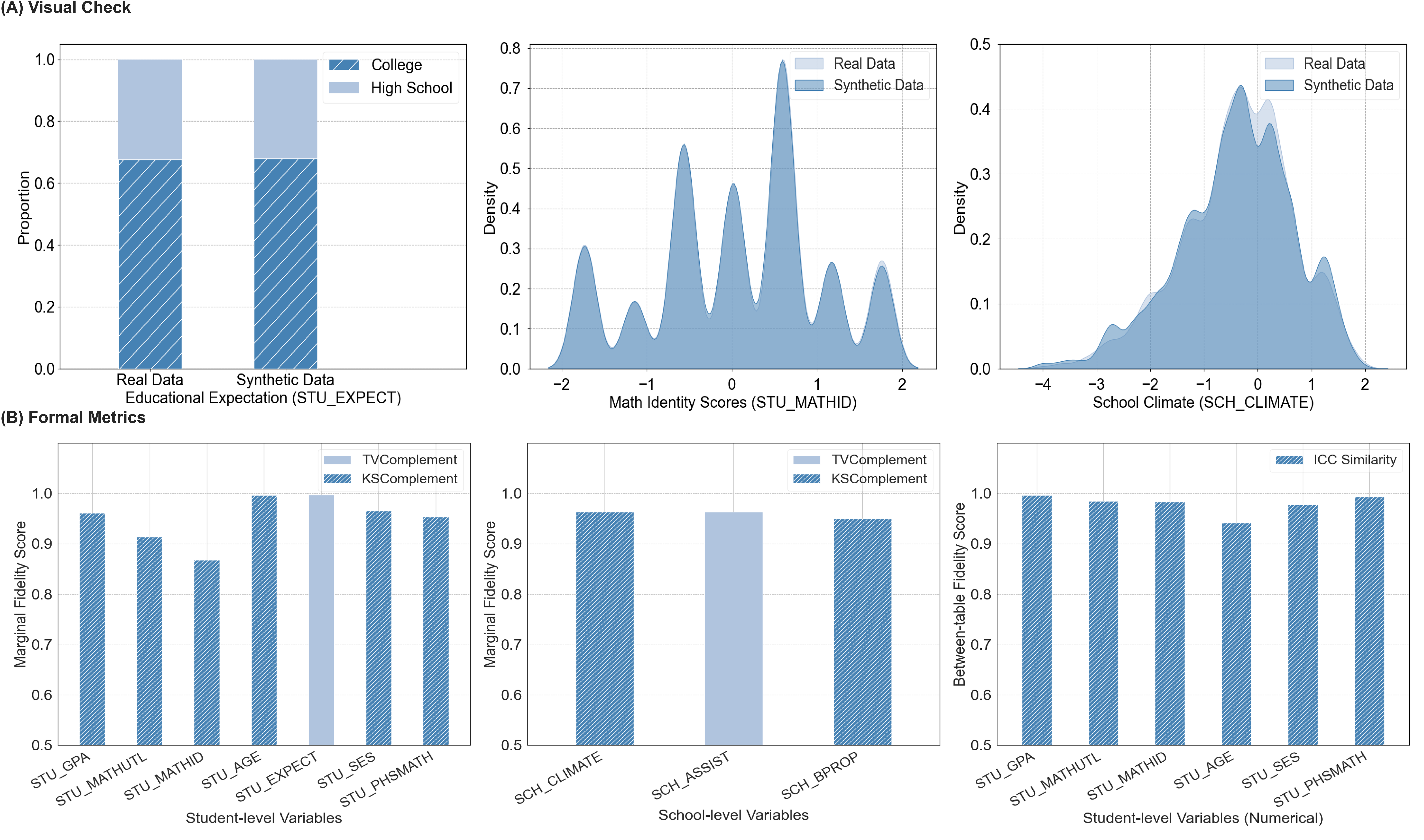}
    \caption{Quality results using visual checks (top) and formal metrics (bottom) on synthetic data generated by ClavaDDPM. TVC = Total Variation Distance complement; KSC = Kolmogorov-Smirnov complement; ICC = Intra-class correlation.} 
    \label{fig:quality_formal_demo2}
\end{figure}

Figure~\ref{fig:quality_formal_demo2} summarizes the data quality results in terms of visual checks and formal metrics. The visual checks indicate that ClavaDDPM-generated synthetic data preserve the marginal distributions from the real data. The formal metrics also exhibit good to excellent between-table fidelity for all variables.

\begin{figure}[!htb]
    \centering
    \includegraphics[width=\linewidth]{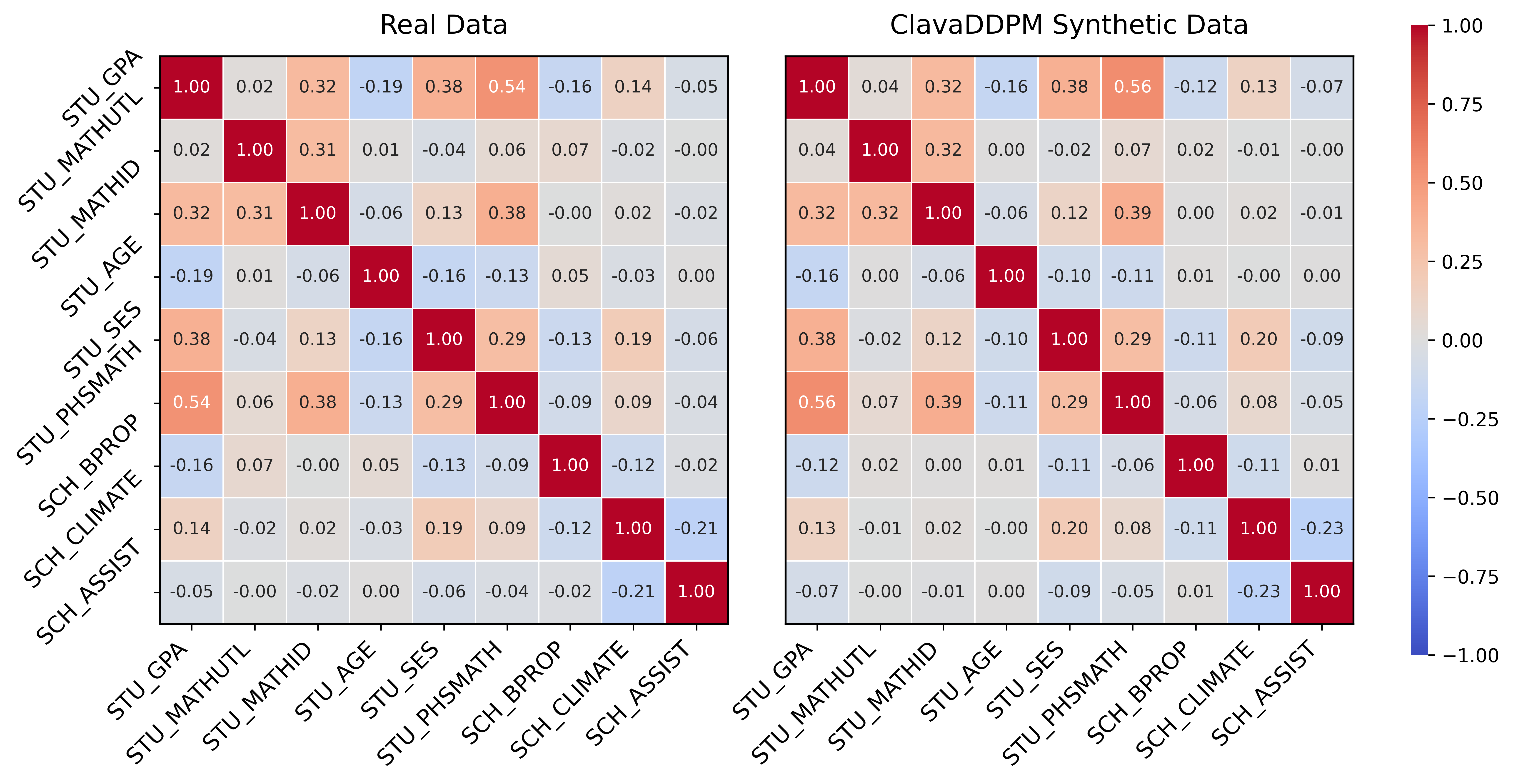}
    \caption{Pearson correlation heatmaps for numerical variables, comparing the real data (left) with the synthetic data generated by ClavaDDPM (right).}
    \label{fig:heat_demo2}
\end{figure}

Figure~\ref{fig:heat_demo2} presents the correlation heatmaps for all numerical variables used in Demonstration 2. ClavaDDPM closely captures pairwise correlations in the real data, with the largest absolute difference between real and synthetic correlations being 0.06. The corresponding correlation similarity scores within the student and school tables are 0.992 and 0.983, respectively, and the 1-hop correlation similarity score is 0.989.

Additionally, the average within-table fidelity scores at the student and school levels are 0.962 and 0.821, respectively, and the average between-table fidelity score is 0.896. ClavaDDPM also achieves an average ML efficacy score of 0.972 for student-level variables and 0.948 for school-level variables. Details of the quality evaluation are provided in Appendix~\ref{app:quality_evaluation}. Overall, the synthetic data generated by ClavaDDPM are of high quality, so we utilize them for our AI-based simulations.

\subsection{Stage 4: Designing a simulation procedure and conducting simulations.} \label{sec:demo2_stage4}

To design the AI-based simulation study for \textcite{huang2017multilevel}, we follow a procedure similar to the one used for predictive performance in Section \ref{sec:demo_step4}. However, we adapt Step 4 specifically for parameter recovery:

\begin{enumerate}
    \item \textit{Set simulation conditions}. Following \textcite{huang2017multilevel}, the main simulation factor of interest is the sample size that varies the number of clusters: $J \in \{10, 30, 50\}$. We also include a condition with $J=100$ to investigate further trends.
    \item \textit{Select performance criteria}. We use the same three performance criteria: the relative bias of point estimates (coefficient bias), the relative bias of standard errors (SE bias), and the coverage probability of 95\% confidence intervals. See \textcite{huang2017multilevel} for detailed equations.
    \item \textit{Select the number of replications}. The number of replications is set to 500 per condition.
    \item \textit{Generate synthetic datasets and outcomes}. Unlike predictive performance, parameter recovery requires known true parameters. First, we determine the true values of the regression coefficients and random-effect components by fitting an assumed HLM model to the empirical HSLS data containing six student-level variables and three school-level variables (e.g., a coefficient value of 0.223 for \texttt{STU\_SES}; a coefficient value of -0.010 for \texttt{SCH\_ASSIST}). Second, using the trained ClavaDDPM model, we generate 500 synthetic datasets for each sample size condition $J \in \{10, 30, 50, 100\}$. Third, we generate a new synthetic outcome for each dataset using the assumed HLM data-generating model, coupled with the data-driven true parameter values.\footnote{We do not directly use the synthetic outcome variable generated by the GenAI model since we need to strictly generate the outcome from the known model in the parameter recovery study.}  
    \item \textit{Apply quantitative methods of interest to synthetic datasets}. We fit the four models of interest---OLS, OLS+IGM, HLM, and HLM+IGM---to each synthetic dataset. To examine the impact of an omitted school-level variable as done in \textcite{huang2017multilevel}, we exclude the \texttt{SCH\_CLIMATE} variable when fitting these four models. We also fit the oracle model that contains all the independent variables as a baseline; see Appendix \ref{app:model_specification} for the model implementation. 
    \item \textit{Measure the performance under the chosen criteria}. We measure the coefficient bias, SE bias, and coverage probability for the coefficient of one student-level variable (\texttt{STU\_SES}) and one school-level variable (\texttt{SCH\_ASSIST}) across all replications for each model. 
\end{enumerate}

For comparison, we replicate the traditional simulation design following the procedure outlined by \textcite{huang2017multilevel}. We generate two school-level predictors from a bivariate normal distribution with correlation $\rho=-0.25$, one of which is treated as unobserved and omitted from the downstream analysis. We then generate one student-level predictor that correlates with the unobserved school-level variable at $\rho=0.25$, where the cluster sizes are unbalanced around an average of 20. Finally, the outcome is generated using a linear function of the three variables with a coefficient set of $(2,2,5)$, a cluster random effect $\mu_j\sim N(0,1)$, and a residual error $r_{ij} \sim N(0,9)$.

\subsection{Stage 5: Evaluating method performance}   

\begin{figure}[!b]
    \centering
    \includegraphics[width=\linewidth]{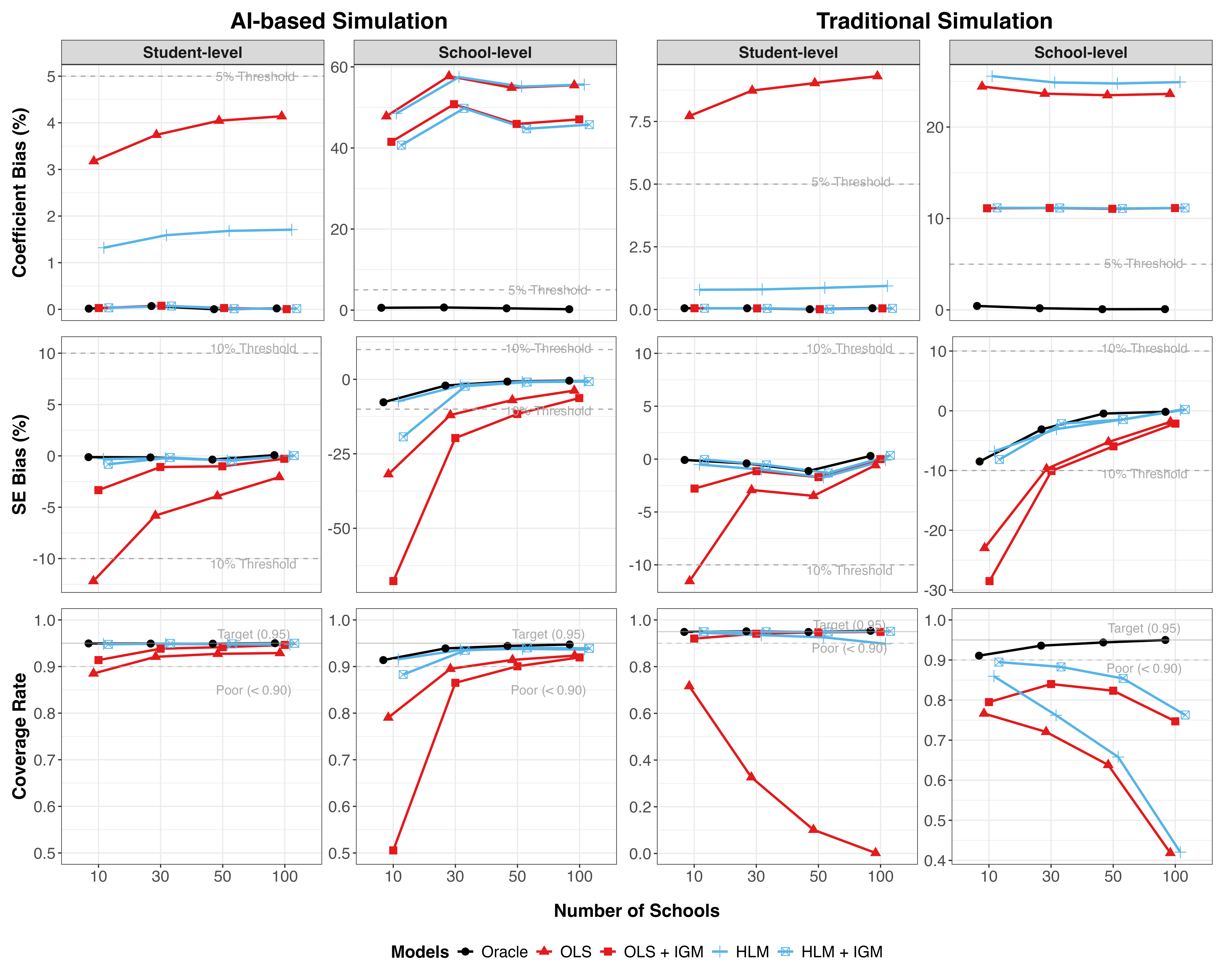}
    \caption{Results of the four models under the AI-based and traditional simulations. Jitter is added to clarify overlapping curves.}
    \label{fig:demo2}
\end{figure}

Figure~\ref{fig:demo2} summarizes the simulation results across three performance criteria: coefficient bias, SE bias, and coverage rate. Jitter is added to clarify overlapping results. The AI-based simulation demonstrates similar trends to the traditional simulation for both coefficient bias (top panels) and SE bias (middle panels). Specifically, regarding student-level coefficient bias, results from both simulations indicate that the OLS and HLM models exhibit biased point estimates, and the bias increases as the number of schools increases. However, the bias of the HLM model is consistently lower than that of OLS. Including the group means of student-level variables nearly eliminates this bias, as the curves of the OLS+IGM and HLM+IGM models overlap with the oracle model. For the school-level point estimates, OLS performs similarly to HLM in both settings, whether the group means of student-level variables are included or not. As seen from the middle panels, the SE bias also shows similar trends between the AI-based and traditional simulations.

However, the coverage rates show different patterns between AI-based and traditional simulations. The AI-based simulation exhibits increasing coverage rates for both student-level and school-level variables as the number of schools increases, whereas the traditional simulation shows decreasing trends. These differences are driven by variations in the true parameter values and the underlying data complexity. The AI-based simulation uses true parameter values learned from the HSLS data and naturally accounts for the real-world complexities among variables. In contrast, the traditional simulation relies on the researcher-dependent parameter values and uses a smaller number of predictors, which is unrealistic.\footnote{As additional simulations, we modified the traditional simulation design by changing the coefficients and correlations to be similar to those from the observed data. Under the updated simulations, we find that coverage rates increase with an increasing number of schools, and the magnitudes of the coefficient and SE biases are similar to those in the AI-based simulation; see Appendix~\ref{app:demo2_huang_sim2} for details.} Through this demonstration, we confirm that our proposed AI-based simulation approach provides more honest and accurate evaluations of how quantitative methods perform in the real world, compared to traditional simulations based on artificial, researcher-determined scenarios.

\section{Robustness Check: Ablation Study}\label{sec:alation} 

As an optional sixth stage, we assess the robustness of our AI-based simulation studies. This type of robustness check is often referred to as an \textit{ablation study} in the computer science literature \parencite{newell1975tutorial}. In this section, we investigate whether modifying a specific step of the data-generation process, while keeping all other steps the same, affects downstream outcomes with respect to synthetic data quality and simulation outcomes \parencite{suk2025using}. We explore three alternative scenarios across the two demonstrations. First, we train ClavaDDPM without using the variable decomposition strategy. Second, we train ClavaDDPM on a reduced input dataset by randomly selecting half of the schools. Third, we reduce the value of the hyperparameter that controls the number of latent groups from 50 to 20 during ClavaDDPM training.

\subsection{Synthetic Data Quality}

Table~\ref{tab:ablation_quality} summarizes the ablation results on synthetic data quality under the three scenarios, where the original results are provided for comparison. Scenario 1 (no variable decomposition) achieves a slightly higher student-level within-table fidelity (by <1\%) compared to the original baseline across both demonstrations. However, it decreases school-level within-table fidelity by 2-5\%  and between-table fidelity by 3-9\%. In contrast, Scenario 2 (half of the schools) produces a 6-7\% reduction in school-level within-table fidelity. This reduction is expected as ClavaDDPM is trained on a much smaller sample of schools. Furthermore, Scenario 3 (reduced latent groups) reduces school-level fidelity by 2-5\%. Finally, ML efficacy scores remain robust across all three ablation scenarios, with all changes lower than 2\%.

\begin{table}[!ht]
\caption{Robustness checks on synthetic data quality}
\label{tab:ablation_quality}
\begin{threeparttable}
\small
\setlength{\tabcolsep}{1.5pt} 
\begin{tabular}{llcccccccc}
\toprule
 &  & \multicolumn{2}{c}{Original} & \multicolumn{2}{c}{Scenario 1} & \multicolumn{2}{c}{Scenario 2} & \multicolumn{2}{c}{Scenario 3} \\
\cmidrule(lr){3-4} \cmidrule(lr){5-6} \cmidrule(lr){7-8} \cmidrule(lr){9-10}
Domain & Level & Demo 1 & Demo 2 & Demo 1 & Demo 2 & Demo 1 & Demo 2 & Demo 1 & Demo 2 \\
\midrule
\multirow[t]{2}{*}{Within-table fidelity}
  & Student & 0.951 & 0.962 & 0.962 & 0.967 & 0.949 & 0.961 & 0.954 & 0.961  \\
  & School  & 0.966 & 0.821 & 0.910 & 0.805 & 0.893 & 0.766 & 0.920 & 0.805  \\
Between-table fidelity
  &     & 0.901 & 0.896 & 0.815 & 0.862 & 0.894 & 0.870 & 0.903 & 0.884  \\
\multirow[t]{2}{*}{ML efficacy}
  & Student & 0.963 & 0.972 & 0.962 & 0.976  & 0.965 & 0.967  & 0.966 &  0.971 \\
  & School  & 0.825 & 0.948 & 0.847 & 0.959 & 0.812 & 0.956 & 0.838 & 0.962  \\
\bottomrule
\end{tabular}
\vspace{1mm}
\begin{tablenotes}[flushleft]
\footnotesize
\item \textit{Note.} Scenario 1 = no variable decomposition; Scenario 2 = half schools in input data; Scenario 3 = reduced latent groups; Demo 1 = Demonstration 1 (predictive performance); Demo 2 = Demonstration 2 (parameter recovery); ML = machine learning. All metric scores are averaged within each level of each domain.
\end{tablenotes}
\end{threeparttable}
\end{table}

\subsection{Simulation Outcomes}

Table~\ref{tab:ablation_prediction} presents the ablation results for PMSE in Demonstration 1 when the number of schools is 50.\footnote{We observe similar result patterns when the number of schools is set to different values (e.g., $J=10$ or $300$.} In Scenario 1, we find an approximate 4\% increase in PMSE for the OLS$^{\rm FE}$ and prior methods, but a 10\% increase for the multilevel method. This is expected because this scenario does not preserve the clustering effect from the real data. In such a case, there is no additional benefit from using the multilevel method, so its predictions are similar to those from the other two methods. In contrast, simulation results under Scenarios 2 and 3 remain relatively robust. This stability is explained in part by the fact that we only use a small number of school-level predictors in the simulations, thereby making their overall impact minimal.

\begin{table}[!ht]
\caption{Robustness checks on simulation outcomes in Demonstration 1}
\label{tab:ablation_prediction}
\begin{threeparttable}
\small
\setlength{\tabcolsep}{9pt} 
\begin{tabular}{lcccccccc}
\toprule
 &  \multicolumn{2}{c}{Original} & \multicolumn{2}{c}{Scenario 1} & \multicolumn{2}{c}{Scenario 2} & \multicolumn{2}{c}{Scenario 3} \\
\cmidrule(lr){2-3} \cmidrule(lr){4-5} \cmidrule(lr){6-7} \cmidrule(lr){8-9}
Method & PMSE & (SE) & PMSE & (SE) & PMSE & (SE) & PMSE & (SE) \\
\midrule
OLS$^{\rm FE}$
  &  55.80 & (5.66) & 59.25 & (5.76) & 56.22  & (5.58) & 55.62 & (5.97)  \\
Prior
  &  55.66 & (5.73) & 59.39 & (5.80) & 55.84 & (5.62) & 55.37 & (5.91)  \\
Multilevel
  & 50.83 & (5.15) & 59.49 & (5.76) & 51.27 & (5.22) & 50.61 & (5.31)  \\
\bottomrule
\end{tabular}
\vspace{1mm}
\begin{tablenotes}[flushleft]
\footnotesize
\item \textit{Note.} Scenario 1 = no variable decomposition; Scenario 2 = half schools in input data; Scenario 3 = reduced latent groups; PMSE = prediction mean squared error; OLS$^{\rm FE}$ = ordinary least squares with school fixed intercepts (and if applicable, school fixed slopes); Prior = multilevel prediction ignoring school-specific random effects; Multilevel = multilevel prediction from the hierarchical linear model.
\end{tablenotes}
\end{threeparttable}
\end{table}

Table~\ref{tab:ablation_parameter} summarizes the ablation results for the simulation outcomes of the OLS+IGM model in Demonstration 2 when the number of clusters is 100. The differences in coefficient bias and SE bias of the student-level coefficient between each ablation scenario and the original are minimal (less than 1\%). In contrast, the school-level coefficient bias shows larger differences of about 3-12\%. In particular, Scenario 2 (half of the schools) yields an approximately 12\% higher coefficient bias. Regarding the 95\% coverage rate, all ablation cases produce results similar to the original baseline. 

\begin{table}[!ht]
\caption{Robustness checks on simulation outcomes in Demonstration 2}
\label{tab:ablation_parameter}
\begin{threeparttable}
\small
\setlength{\tabcolsep}{12pt} 
\begin{tabular}{llrrrr}
\toprule
Criterion  & Level  & Original & Scenario 1 & Scenario 2 & Scenario 3 \\
\midrule
\multirow[t]{2}{*}{Coefficient bias (\%)}
  & Student &  0.004  & 0.009 & 0.015 & 0.020  \\
  & School  & 47.049 & 49.908 & 59.455  & 45.609 \\
\multirow[t]{2}{*}{SE bias (\%)}
  & Student & -0.282 & -0.476 & -0.203 & -0.208  \\
  & School  & -6.293 & -5.898 & -6.460 & -6.323  \\
\multirow[t]{2}{*}{95\% coverage}
  & Student & 0.946  & 0.946  & 0.947  & 0.947  \\
  & School  & 0.920   & 0.919   & 0.914   & 0.920  \\
\bottomrule
\end{tabular}
\vspace{1mm}
\begin{tablenotes}[flushleft]
\footnotesize
\item \textit{Note.} Scenario 1 = no variable decomposition; Scenario 2 = half schools in input data; Scenario 3 = reduced latent groups; SE = standard error.
\end{tablenotes}
\end{threeparttable}
\end{table}

In conclusion, these ablation studies emphasize the importance of optimizing GenAI models to produce high-quality synthetic data for AI-based simulation studies. Without such optimization, the resulting synthetic data are of poorer quality and they yield less accurate evaluations of the real-world performance of the target quantitative methods. Ultimately, unsuccessful data generation undermines the primary goal of designing and executing AI-based simulations.

\section{Discussion and Conclusions}\label{sec:con}

In this paper, we propose a general AI-based simulation approach for evaluating the predictive performance and parameter recovery of statistical methods. Our approach consists of six main stages: (i) specifying a statistical method and real data scheme, (ii) training GenAI models on real data, (iii) assessing synthetic data quality, (iv) designing and conducting simulations, (v) evaluating method performance, and (vi) performing optional robustness checks. We enhance the capabilities of existing GenAI models, namely ClavaDDPM and CTGAN, through proper modifications that help generate realistic multilevel data. Additionally, we provide a formal framework for evaluating the quality of synthetic multilevel data. We also demonstrate the utility of our approach through two ``case studies'' based on prior research: one for predictive performance \parencite{afshartous2005prediction}  and the other  for parameter recovery \parencite{huang2017multilevel}. In the latter case, we illustrate how AI-generated synthetic data can be manipulated to deviate from real data in a controlled way during the simulation design stage. 

The proposed AI-based simulation approach has broad applicability in statistics, methodology, and applied research by assessing old and newly proposed quantitative methods using realistic data.  It can serve as a complementary validation tool or a primary evaluation method. For example, most statistical and methodological research has traditionally evaluated finite-sample performance in ``ideal'' data settings, which are often characterized by normally distributed variables, limited dependencies, or linear relationships. While these traditional simulations could evaluate the finite-sample performance of methods, they ignore the complexity of real-world finite samples. Therefore, our AI-based simulation approach offers an effective complement by evaluating methods in realistic settings. Additionally, providing practical guidance on quantitative methods for applied research is of crucial value, such as the standards by the What Works Clearinghouse (WWC). Developing high-quality guidance requires testing methods on realistic datasets derived from benchmark real data. In this case, our approach offers a flexible and efficient tool for honest method evaluation in real-world scenarios.

Lastly, there are some limitations and suggestions for future research. First, we introduced modifications to ClavaDDPM and CTGAN within their current architectures. Although variable decomposition for ClavaDDPM effectively preserves the ICC of numeric variables, it does not demonstrate superior performance for categorical variables. Similarly, while CLCA restores logical consistency of school-level variables and improves distributional alignment in CTGAN-generated data, it does not fully recover the hierarchical dependence structure of the real data, as reflected in ICC-based measures. Therefore, future research would explore additional architectural modifications or develop new GenAI models explicitly designed to respect the unique characteristics of multilevel data. Second, while our approach addresses predictive performance and parameter recovery, it does not extend to causal effect studies. Simulation studies for causal inference require known causal relationships among variables prior to data generation, and thus, multiple generators may be required (e.g., separate generators for covariates, treatment variable, and outcome variable). Although \textcite{athey2024using} offers an initial proposal on this topic, there are ample opportunities to develop generator architectures and simulation designs tailored for causal inference.

\printbibliography

\newpage 

\setcounter{equation}{0}
\setcounter{figure}{0}
\setcounter{table}{0}
\setcounter{page}{1}
\setcounter{section}{0}
\renewcommand{\theequation}{S\arabic{equation}}
\renewcommand{\thefigure}{S\arabic{figure}}
\renewcommand{\thesection}{S\arabic{section}}
\renewcommand{\thetable}{S\arabic{table}}
\renewcommand{\thefigure}{S\arabic{figure}}
\setlength{\parindent}{30pt}

\setcounter{secnumdepth}{2}

\begin{center}
{\Large
\textbf{Supplementary Materials}
}
\end{center}

\section{Cluster-Level Consistency Alignment (CLCA) Algorithms}\label{app:cc}

We develop the CLCA algorithm for post-processing school-level variables in CTGAN-generated multilevel data: Algorithm \ref{alg:school-numeric-ot} for numeric variables and Algorithm \ref{alg:school-cat-opt} for categorical variables. At a high level, Algorithm~\ref{alg:school-numeric-ot} first computes a school-level summary for each synthetic school, ranks these summaries, and then maps them to the corresponding quantiles of the real school-level distribution. The adjusted value is subsequently assigned to all students within the same school, thereby enforcing within-school consistency while preserving the empirical school-level marginal distribution. Because one-dimensional optimal transport reduces to monotone quantile matching, this procedure implements a rank-based quantile matching procedure.

\begin{algorithm}[H]
\caption{School-Level Numeric Enforcement via Rank-Based Quantile Matching}
\label{alg:school-numeric-ot}
\begin{algorithmic}[1]
\Require Synthetic dataset $\mathbf{D}^{(s)}$, real dataset $\mathbf{D}^{(r)}$, school-level numeric variable $v$
\Ensure Adjusted variable $v'$ where all students in the same school share a common value, and the school-level distribution matches that of the real data

\Statex \textit{Step 1: Compute synthetic school-level medians}
\For{each school $j$ in $\mathbf{D}^{(s)}$}
    \State $\tilde{s}_j \gets \text{median}(v \mid \text{school } j,\, \mathbf{D}^{(s)})$
\EndFor

\Statex
\Statex \textit{Step 2: Quantile matching (1-Dimensional optimal transport)}
\State Let $J \gets |\{\tilde{s}_j\}|$ \Comment{number of synthetic schools}
\State $\pi \gets \text{argsort}(\tilde{s}_1, \dots, \tilde{s}_J)$ \Comment{rank synthetic school medians}
\For{$m = 1, \dots, J$}
    \State $q_m \gets (m - 0.5)\,/\,J$
    \State $\tilde{s}_{\pi(m)}' \gets Q_r(q_m)$ \Comment{$Q_r$: quantile function of real school-level values}
\EndFor

\Statex
\Statex \textit{Step 3: Apply school-level values to all students in $\mathbf{D}^{(s)}$}
\For{each school $j$ in $\mathbf{D}^{(s)}$}
    \State Set $v' \gets \tilde{s}_j'$ for all students in school $j$
\EndFor
\State \Return $v'$
\end{algorithmic}
\end{algorithm}

\newpage

At a high level, Algorithm~\ref{alg:school-cat-opt} assigns a single category to each synthetic school through a binary optimization problem. The objective balances fidelity to the real student-level marginals with fidelity to the real school-level category counts, while enforcing consistency of school-level variables within each school. The assigned category is then copied to all students within the same school, thereby restoring within- school consistency. Specifically, the CTGAN confidence cost favors school-level assignments that are already supported by the CTGAN confidence signals $\hat{p}_{j,k}$. If $\hat p_{j,k}$ is large, then $-\log \hat p_{j,k}$ is small, so assigning school $j$ to category $k$ incurs a low penalty; if $\hat p_{j,k}$ is small, the penalty is large. We set $\alpha=\beta=\gamma=1$, and replace $\hat p_{j,k}$ by $\max(\hat p_{j,k}, \varepsilon)$ with $\varepsilon=10^{-6}$ to avoid undefined logarithmic costs when a within-school CTGAN proportion is zero.

\noindent\resizebox{1\textwidth}{!}{%
\begin{minipage}{1.1\textwidth}
\begin{algorithm}[H]
\caption{School-Level Categorical Enforcement via Constrained Optimization}
\label{alg:school-cat-opt}
\begin{algorithmic}[1]
\Require Synthetic dataset $\mathbf{D}^{(s)}$, real dataset $\mathbf{D}^{(r)}$, school-level categorical variable $v$ with categories $\{c_1, \dots, c_K\}$
\Require Weights $\alpha, \beta, \gamma \geq 0$
\Ensure Adjusted variable $v'$ where all students in the same school share a common category, and the overall distribution matches that of the real data

\Statex \textit{Step 1: Compute targets from real data}
\State $\mathbf{p}^* \gets \text{prop}(v, \mathbf{D}^{(r)})$ \Comment{student-level proportions per category}
\State $\mathbf{t}^* \gets \text{count}(v, \mathbf{D}^{(r)})$ \Comment{number of schools per category}

\Statex
\Statex \textit{Step 2: Extract CTGAN confidence signals}
\For{each school $j$ in $\mathbf{D}^{(s)}$}
    \State $\hat{\mathbf{p}}_j \gets \text{prop}(v \mid \text{school } j,\, \mathbf{D}^{(s)})$ \Comment{$\hat{p}_{j,k}$: proportion of category $c_k$ in school $j$}
\EndFor

\Statex
\Statex \textit{Step 3: Optimize category selection for each school}
\State \textbf{Decision variables:} $x_{j,k} \in \{0, 1\}$, where $x_{j,k} = 1$ iff school $j$ is assigned category $c_k$
\State \textbf{Constraint:} $\sum_{k=1}^{K} x_{j,k} = 1 \quad \forall\, j$ \Comment{each school selects exactly one category}

\Statex
\State \textbf{Minimize:}
\[
\alpha \underbrace{\sum_{k=1}^{K} \left| \frac{1}{N}\sum_{j} n_j\, x_{j,k} - p_k^* \right|}_{\text{(A) student marginal deviation}}
+ \;\beta \underbrace{\sum_{k=1}^{K} \left| \sum_{j} x_{j,k} - t_k^* \right|}_{\text{(B) school count deviation}}
+ \;\gamma \underbrace{\sum_{j}\sum_{k=1}^{K} \bigl(-\log \hat{p}_{j,k}\bigr)\, x_{j,k}}_{\text{(C) CTGAN confidence cost}}
\]
\Statex where $n_j$ is the number of students in school $j$ and $N = \sum_j n_j$

\Statex
\Statex \textit{Step 4: Apply school-level assignments to all students in $\mathbf{D}^{(s)}$}
\For{each school $j$ in $\mathbf{D}^{(s)}$}
    \State Set $v' \gets c_k$ where $x_{j,k} = 1$ for all students in school $j$
\EndFor
\State \Return $v'$
\end{algorithmic}
\end{algorithm}
\end{minipage}}

\newpage

\section{HSLS Data Schema}\label{app:HSLSdata}  

The dataset used to train ClavaDDPM follows a two-level, multi-table structure in which students are nested within schools. As shown in Figure~\ref{fig:data_schema}, the school table and the student table are linked through \texttt{SCH\_ID}, which serves as the primary key in the school table and the foreign key in the student table. Depending on the demonstration, we use different sets of variables. 

\begin{figure}[ht]
\centering

\subcaptionbox{Demonstration 1: Predictive Performance\label{fig:schema_prediction}}[\textwidth]{%
\resizebox{\textwidth}{!}{%
  \begin{tikzpicture}[>=latex, every node/.style={transform shape}]
  \matrix (school) [
      matrix of nodes,
      nodes in empty cells,
      nodes={font=\footnotesize, minimum height=4mm, inner ysep=0.5pt},
      column sep=-\pgflinewidth,
      row sep=-\pgflinewidth,
      column 1/.style={anchor=west,  minimum width=3.2cm},
      column 2/.style={anchor=west, minimum width=3.0cm},
      ampersand replacement=\&
  ]{
    SCH\_ID      \& Primary Key   \\
    SCH\_BPROP   \& Numerical     \\
    SCH\_LOCALE  \& Categorical   \\
  };
  \node[anchor=south west] at ($(school-1-1.north west)+(0,2mm)$) {\textbf{School Table}};
  \draw (school-1-1.north west) rectangle (school-3-2.south east);
  \draw (school-1-1.north east) -- (school-3-1.south east);
  \draw (school-1-1.south west) -- (school-1-2.south east);
  \matrix (student) [
      matrix of nodes,
      nodes in empty cells,
      nodes={font=\footnotesize, minimum height=4mm, inner ysep=0.5pt},
      column sep=-\pgflinewidth,
      row sep=-\pgflinewidth,
      right=3cm of school,
      column 1/.style={anchor=west,  minimum width=4.2cm},
      column 2/.style={anchor=west, minimum width=3.2cm},
      ampersand replacement=\&
  ]{
    STU\_ID          \& Primary Key   \\
    SCH\_ID          \& Foreign Key   \\
    STU\_SMATH        \& Numerical     \\
    STU\_MATHUTL     \& Numerical     \\
    STU\_MATHID      \& Numerical     \\
    STU\_AGE         \& Numerical     \\
    STU\_SES         \& Numerical     \\
    STU\_PHSMATH     \& Numerical     \\
    STU\_RACE        \& Categorical   \\
    STU\_EXPECT      \& Categorical   \\
  };
  \node[anchor=south west] at ($(student-1-1.north west)+(0,2mm)$) {\textbf{Student Table}};
  \draw (student-1-1.north west) rectangle (student-10-2.south east);
  \draw (student-1-1.north east) -- (student-10-1.south east);
  \draw (student-2-1.south west) -- (student-2-2.south east);
  \draw[<->] (school-1-2.east) -- (student-2-1.west);
  \end{tikzpicture}%
}%
}

\vspace{6mm}

\subcaptionbox{Demonstration 2: Parameter Recovery\label{fig:schema_recovery}}[\textwidth]{%
\resizebox{\textwidth}{!}{%
  \begin{tikzpicture}[>=latex, every node/.style={transform shape}]
  \matrix (school) [
      matrix of nodes,
      nodes in empty cells,
      nodes={font=\footnotesize, minimum height=4mm, inner ysep=0.5pt},
      column sep=-\pgflinewidth,
      row sep=-\pgflinewidth,
      column 1/.style={anchor=west,  minimum width=3.2cm},
      column 2/.style={anchor=west, minimum width=3.0cm},
      ampersand replacement=\&
  ]{
    SCH\_ID       \& Primary Key   \\
    SCH\_BPROP    \& Numerical     \\
    SCH\_CLIMATE  \& Numerical     \\
    SCH\_ASSIST  \& Numerical     \\
  };
  \node[anchor=south west] at ($(school-1-1.north west)+(0,2mm)$) {\textbf{School Table}};
  \draw (school-1-1.north west) rectangle (school-4-2.south east);
  \draw (school-1-1.north east) -- (school-4-1.south east);
  \draw (school-1-1.south west) -- (school-1-2.south east);
  \matrix (student) [
      matrix of nodes,
      nodes in empty cells,
      nodes={font=\footnotesize, minimum height=4mm, inner ysep=0.5pt},
      column sep=-\pgflinewidth,
      row sep=-\pgflinewidth,
      right=3cm of school,
      column 1/.style={anchor=west,  minimum width=4.2cm},
      column 2/.style={anchor=west, minimum width=3.2cm},
      ampersand replacement=\&
  ]{
    STU\_ID          \& Primary Key   \\
    SCH\_ID          \& Foreign Key   \\
    STU\_GPA         \& Numerical     \\
    STU\_MATHUTL     \& Numerical     \\
    STU\_MATHID      \& Numerical     \\
    STU\_AGE         \& Numerical     \\
    STU\_SES         \& Numerical     \\
    STU\_PHSMATH     \& Numerical     \\
    STU\_EXPECT      \& Categorical   \\
  };
  \node[anchor=south west] at ($(student-1-1.north west)+(0,2mm)$) {\textbf{Student Table}};
  \draw (student-1-1.north west) rectangle (student-9-2.south east);
  \draw (student-1-1.north east) -- (student-9-1.south east);
  \draw (student-2-1.south west) -- (student-2-2.south east);
  \draw[<->] (school-1-2.east) -- (student-2-1.west);
  \end{tikzpicture}%
}%
}

\caption{Data Schema of the School and Student Tables for Demonstration 1 (a) and Demonstration 2 (b)}
\label{fig:data_schema}
\end{figure}

Table~\ref{tab:variables} shows all variables used in the two demonstrations and their descriptions. 
\begin{table}[!ht]
    \centering
    \caption{Variables used in Demonstrations 1 and 2}
    \label{tab:variables}
    \begin{threeparttable}
    \begin{tabularx}{\textwidth}{@{}lXl}
        \toprule
        \textbf{Variables} & \textbf{Description} \\
        \midrule
        \texttt{STU\_ID} & Student's ID. \\
        \texttt{SCH\_ID} & School's ID.  \\        
        \texttt{STU\_SMATH} & Standardized math achievement test scores.\\
        \texttt{STU\_GPA} & Academic GPA at the completion of Grade 9. \\
        \texttt{STU\_MATHUTL} & A psychometric scale measuring students' perceived utility of mathematics; higher scores indicate a greater perceived value.  \\
        \texttt{STU\_MATHID} & A psychometric scale measuring the strength of a student's mathematical identity.  \\
        \texttt{STU\_AGE} & Student's age  \\
        \texttt{STU\_EXPECT} & Student's educational expectations (1 = College degree or higher; 0 = High school diploma or less).  \\
        \texttt{STU\_SES} & Socioeconomic status (SES) index.  \\
        \texttt{STU\_PHSMATH} & Math achievement score achieved prior to entering high school.  \\
        \texttt{STU\_RACE} & Race/ethnicity.  \\

        \texttt{SCH\_BPROP} & School's proportion of Black students.  \\
        \texttt{SCH\_LOCALE} & The school's geographic setting (e.g., city, suburb, town, or rural).  \\
        \texttt{SCH\_CLIMATE} & A measure of the school's social and behavioral environment; higher scores indicate fewer reported school-wide problems. \\
        \texttt{SCH\_ASSIST} & Level of assistance provided by school counselors during the student's transition into high school.  \\
        \bottomrule
    \end{tabularx}
    \vspace{1ex}
    \begin{tablenotes}
        \small
        \item \textit{Note.} Variables beginning with ``\texttt{STU\_}'' and ``\texttt{SCH\_}'' denote student-level and school-level variables, respectively.
    \end{tablenotes}
    \end{threeparttable}
\end{table}

\vspace{\fill}\clearpage

\section{Model Specification}\label{app:model_specification}

In Demonstration 1, following \textcite{afshartous2005prediction}, we consider three prediction rules. The multilevel model of interest is a \textit{random slope} model, specified as follows using the \texttt{lmer} function from the R package \texttt{lme4} \parencite{Bates2015Fitting}:
\begin{equation*}
\begin{aligned}
\texttt{STU\_SMATH} &\sim \texttt{STU\_MATHUTL} + \texttt{STU\_MATHID} + \texttt{STU\_AGE} \\
&\quad + \texttt{STU\_EXPECT} + \texttt{STU\_SES} + \texttt{STU\_RACE} \\
&\quad + \texttt{STU\_PHSMATH} + \texttt{SCH\_LOCALE} + \texttt{SCH\_BPROP} \\
&\quad + (1 + \texttt{STU\_MATHUTL} \mid \texttt{SCH\_ID}) 
\end{aligned}.
\end{equation*}
This multilevel model uses standardized math performance as the outcome and includes selected student-level and school-level predictors, as well as school-specific random intercepts and a random slope for \texttt{STU\_MATHUTL}. We include the random slope for \texttt{STU\_MATHUTL} to account for its potential cluster-specific varying effect. We do not estimate random slopes for all student-level predictors due to insufficient school-level variability. In contrast, \textcite{afshartous2005prediction} used an idealized simulation design with only one student-level predictor and one school-level predictor. The school-level predictor was used to predict both the intercept and the slope of the student-level predictor. The \textit{multilevel prediction rule} uses all the fixed-effects and random-effects components of the multilevel model, whereas the \textit{prior prediction rule} uses only the fixed-effects components. 

Even with only one random slope, we sometimes face statistical fitting issues (e.g., negative eigenvalues), in particular under small sample size conditions (e.g., 10 clusters). In such cases, we reduce the model to a random-intercept-only model. This fallback preserves the multilevel structure of the model while avoiding overparameterization.

The OLS model with school fixed effects (OLS$^{\rm FE}$) in Demonstration 1 is specified as:
\begin{equation*}
\begin{aligned}
\texttt{STU\_SMATH} &\sim \texttt{STU\_MATHUTL} + \texttt{STU\_MATHID} + \texttt{STU\_AGE} \\
&\quad + \texttt{STU\_EXPECT} + \texttt{STU\_SES} + \texttt{STU\_RACE} \\
&\quad + \texttt{STU\_PHSMATH} + \texttt{SCH\_ID} + \texttt{STU\_MATHUTL}:\texttt{SCH\_ID}  \\
\end{aligned}.
\end{equation*}
The \textit{OLS prediction rule} is based on this OLS$^{\rm FE}$ model, which represents a realistic fixed-effects model that can be fitted to the empirical HSLS dataset. However, \textcite{afshartous2005prediction} allows for school-specific fixed intercepts and school-specific fixed slopes for the one student-level predictor in the idealized traditional simulation design.

In Demonstration 2, based on \textcite{huang2017multilevel}, we assume the following HLM with school-specific random intercepts to generate the new synthetic outcome $Y$:  
\begin{equation*}
Y = \beta_0 + \sum_{p=1}^{6} \beta_{\text{STU}, p} X_{\rm STU} + \sum_{q=1}^{3} \beta_{\text{SCH}, q} X_{\rm SCH} + U_j + R_{ij}, \quad  U_j \sim N(0,\sigma_U^2), \quad{} R_{ij} \sim N(0,\sigma_R^2).
\end{equation*}
The student-level predictors $X_{\rm STU}$ include  \texttt{STU\_MATHUTL},  \texttt{STU\_MATHID}, \texttt{STU\_AGE}, \texttt{STU\_EXPECT}, \texttt{STU\_SES}, and \texttt{STU\_PHSMATH}. The student-level predictors $X_{\rm SCH}$ include \texttt{SCH\_BPROP}, \texttt{SCH\_ASSIST}, and \texttt{SCH\_CLIMATE}. 

Following \textcite{huang2017multilevel}, we fit four models: (i) OLS, (ii) OLS including group means of student-level variables (OLS+IGM), (iii) HLM, and (iv) HLM including group means of student-level variables (HLM+IGM). To study the effect of omitted variable bias, we intentionally exclude the \texttt{SCH\_CLIMATE} variable, which is significantly correlated with \texttt{STU\_SES} ($\rho = 0.19$) and \texttt{SCH\_ASSIST}($\rho=-0.21$), when fitting models. Specifically, the OLS model is specified as:
\begin{equation*}
\begin{aligned}
\texttt{STU\_GPA} &\sim \texttt{STU\_MATHUTL} + \texttt{STU\_MATHID} + \texttt{STU\_AGE} \\
&\quad + \texttt{STU\_EXPECT} + \texttt{STU\_SES} + \texttt{STU\_PHSMATH} \\
&\quad + \texttt{SCH\_BPROP} + \texttt{SCH\_ASSIST}
\end{aligned}
\end{equation*}
and the OLS+IGM model is specified as: 
\begin{equation*}
\begin{aligned}
\texttt{STU\_GPA} &\sim \texttt{STU\_MATHUTL} + \texttt{STU\_MATHID} + \texttt{STU\_AGE} \\
&\quad + \texttt{STU\_EXPECT} + \texttt{STU\_SES} + \texttt{STU\_PHSMATH} \\
&\quad + \texttt{SCH\_BPROP} + \texttt{SCH\_ASSIST} \\
&\quad + \overline{\texttt{STU\_MATHUTL}} + \overline{\texttt{STU\_MATHID}} + \overline{\texttt{STU\_AGE}} \\
&\quad + \overline{\texttt{STU\_EXPECT}} + \overline{\texttt{STU\_SES}} + \overline{\texttt{STU\_PHSMATH}}
\end{aligned}
\end{equation*}
where $\overline{\cdot}$ represents the school mean of a student-level variable. 

Similarly, we specify the HLM and HLM+IGM models as follows:
\begin{equation*}
\begin{aligned}
\texttt{STU\_GPA} &\sim \texttt{STU\_MATHUTL} + \texttt{STU\_MATHID} + \texttt{STU\_AGE} \\
&\quad + \texttt{STU\_EXPECT} + \texttt{STU\_SES} + \texttt{STU\_PHSMATH} \\
&\quad + \texttt{SCH\_BPROP} + \texttt{SCH\_ASSIST}  \\
&\quad + (1 + \mid \texttt{SCH\_ID}) 
\end{aligned}
\end{equation*}
\begin{equation*}
\begin{aligned}
\texttt{STU\_GPA} &\sim \texttt{STU\_MATHUTL} + \texttt{STU\_MATHID} + \texttt{STU\_AGE} \\
&\quad + \texttt{STU\_EXPECT} + \texttt{STU\_SES} + \texttt{STU\_PHSMATH} \\
&\quad + \texttt{SCH\_BPROP} + \texttt{SCH\_ASSIST} \\
&\quad + \overline{\texttt{STU\_MATHUTL}} + \overline{\texttt{STU\_MATHID}} + \overline{\texttt{STU\_AGE}} \\
&\quad + \overline{\texttt{STU\_EXPECT}} + \overline{\texttt{STU\_SES}} + \overline{\texttt{STU\_PHSMATH}} \\
&\quad + (1 + \mid \texttt{SCH\_ID}) 
\end{aligned}
\end{equation*}

\newpage

\section{Quality Evaluation}\label{app:quality_evaluation}

\begin{figure}[!h]
    \centering
    \includegraphics[width=\linewidth]{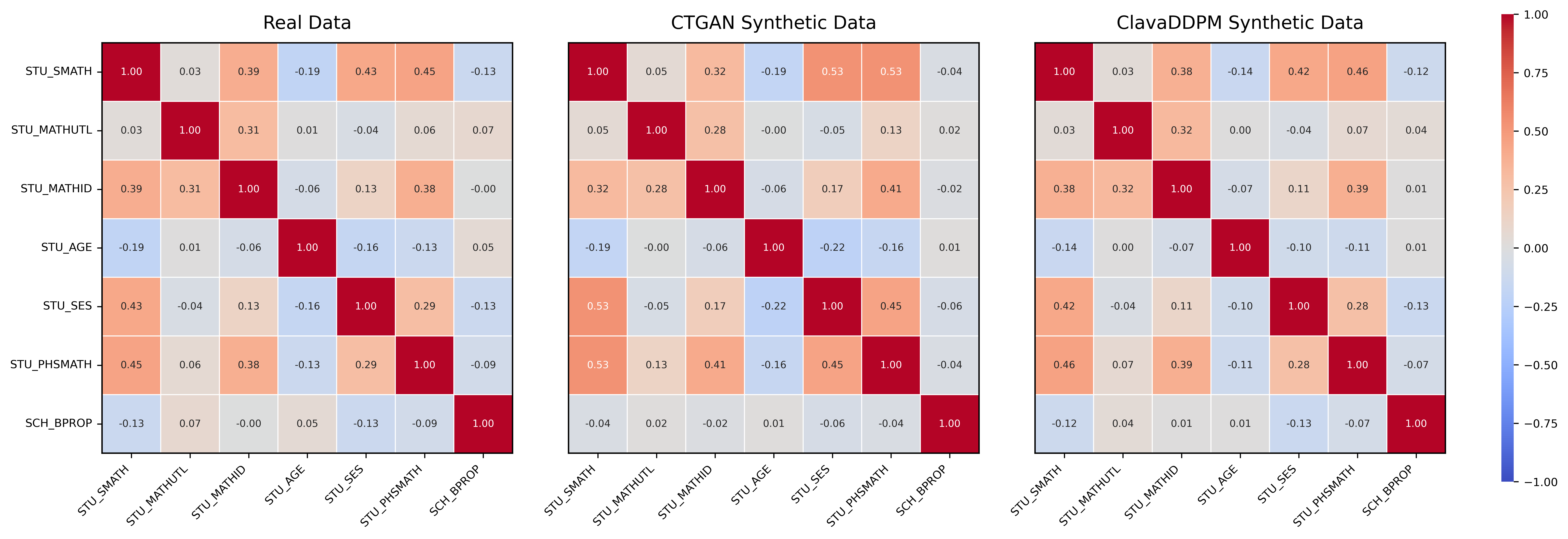}
    \caption{Pearson correlation heatmaps for the variables used in Demonstration 1, comparing the real data (left) against synthetic data generated by CTGAN (middle) and ClavaDDPM (right).}
    \label{fig:corr_map_demo1}
\end{figure}

In this section, we provide two sets of correlation heatmaps and one table of formal metrics. Figure~\ref{fig:corr_map_demo1} compares the correlation matrix of the real data (left panel) with that of the synthetic data generated by CTGAN (middle panel) and ClavaDDPM (right panel) in Demonstration 1. Although both models achieve strong correlation fidelity, ClavaDDPM better preserves these pairwise relationships. For example, \texttt{SCH\_BPROP}'s correlations with other variables differ from those in the real data by less than 0.03, whereas the corresponding differences under CTGAN range from 0.02 to 0.07.

\begin{figure}[H]
    \centering
    \includegraphics[width=\linewidth]{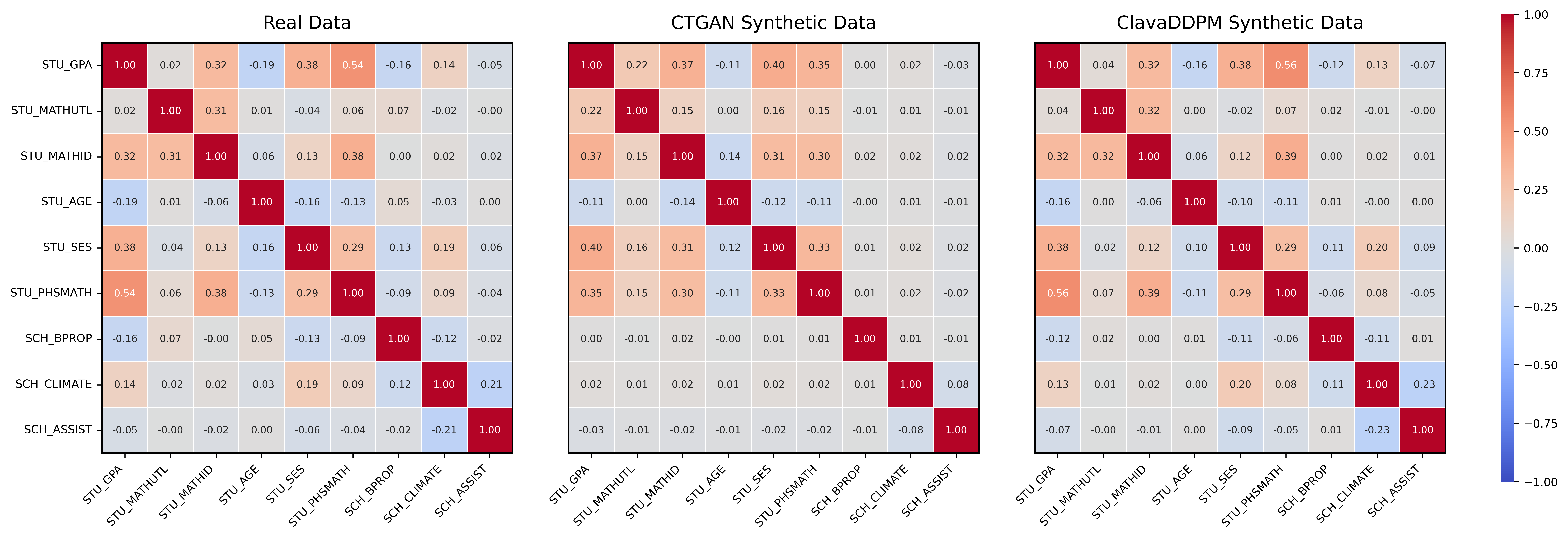}
    \caption{Pearson correlation heatmaps for the variables used in Demonstration 2, comparing the real data (left) against synthetic data generated by CTGAN (middle) and ClavaDDPM (right)}
    \label{fig:corr_map_demo2}
\end{figure}

Similarly, Figure~\ref{fig:corr_map_demo2} presents the correlation heatmap for Demonstration 2, where ClavaDDPM again outperforms CTGAN in correlation fidelity. For the variables \texttt{STU\_PHSMATH}, \texttt{SCH\_BPROP}, \texttt{SCH\_CLIMATE}, and \texttt{STU\_ASSIST}, the absolute differences between real and synthetic correlations are below 0.06 under ClavaDDPM, whereas they range from 0.01 to 0.20 under CTGAN.

Table~\ref{tab:demo12_fidelity} reports the formal metrics of within-table fidelity, between-table fidelity, and ML efficacy for Demonstrations 1 and 2. Both models achieve good to excellent average scores in general, and ClavaDDPM outperforms CTGAN on nearly every metric. 

We remark that during the training stage, we evaluated both ClavaDDPM and CTGAN on two datasets: a subset that contains the selected variables (as shown in Figure~\ref{fig:data_schema}) and the full dataset that contains all 28 variables. We found that ClavaDDPM produces higher-fidelity synthetic data when trained on the subset, whereas CTGAN achieves higher fidelity when trained on the full dataset. Therefore, to optimize performance, we train our final ClavaDDPM model on the subset and our CTGAN model on the full dataset, and then summarize the synthetic data quality from each model.

\begin{table}[!h]
    \centering
    \begin{threeparttable}
    \caption{Formal Metric Evaluation}
    \label{tab:demo12_fidelity}
    \footnotesize
    \setlength{\tabcolsep}{3pt}
    \begin{tabular}{llcccc}
        \toprule
        \multirow{2}{*}{\textbf{Domain}} & \multirow{2}{*}{\textbf{Metric}} & \multicolumn{2}{c}{\textbf{Demonstration 1}} & \multicolumn{2}{c}{\textbf{Demonstration 2}} \\
        \cmidrule(lr){3-4} \cmidrule(lr){5-6}
        & & \textbf{ClavaDDPM} & \textbf{CTGAN} & \textbf{ClavaDDPM} & \textbf{CTGAN} \\
        
        \midrule
        \multicolumn{6}{l}{\textbf{Within-Table Fidelity}} \\
        \multicolumn{6}{l}{\textit{Student Table}} \\
        \multirow[t]{2}{*}{- Marginal} & KSC & 0.946& 0.898&   0.944 & 0.902\\
         & TVC &  0.985 &0.947&  0.992 & 0.967\\
        \multirow[t]{4}{*}{- Pairwise} & Correlation similarity & 0.991 & 0.975 & 0.992  & 0.975 \\
         & Contingency similarity & 0.936 & 0.781 &  0.943 & 0.719 \\
         & Eta-squared similarity & 0.989 & 0.980 &  0.993 & 0.982\\
         & MI similarity & 0.861 & 0.780&  0.910 & 0.759\\
         \multirow[t]{2}{*}{- Average} & Marginal fidelity & 0.966 & 0.923 &  0.968 & 0.935\\
          & Pairwise fidelity & 0.944 & 0.879 &  0.960 & 0.859\\  
          & Student-table fidelity & 0.951 & 0.894 &  0.962 & 0.884\\           
        \multicolumn{6}{l}{\textit{School Table}} \\
        \multirow[t]{2}{*}{- Marginal} & KSC & 0.965 & 0.999 &  0.968 & 0.993\\
         & TVC & 0.976 & 0.963 &  --- & --- \\
        \multirow[t]{4}{*}{- Pairwise} & Correlation similarity & --- & --- & 0.983 & 0.942\\
         & Contingency similarity & 0.927 & 0.916 &  --- & --- \\
         & Eta-squared similarity  & 0.994  & 0.985 & --- & --- \\
         & MI similarity & --- & --- & 0.512 & 0.308\\
         \multirow[t]{2}{*}{- Average} & Marginal fidelity &0.971 &  0.981  & 0.968 &0.993 \\
          & Pairwise fidelity & 0.961 & 0.951 &  0.748 & 0.625\\  
          & School-table fidelity & 0.966 & 0.966 & 0.821 & 0.748\\        
        \multicolumn{6}{l}{\textbf{Between-Table Fidelity}} \\
         & Referential integrity & 1.000 & 1.000 & 1.000 & 1.000 \\
         & Cardinality shape similarity & 0.956 & 0.925 &  0.977 & 0.899 \\
         & k-hop correlation similarity & 0.990 & 0.973&  0.989 & 0.983 \\
         & k-hop contingency similarity & 0.936 & 0.842 & 0.952  & 0.951 \\
         & k-hop Eta-squared similarity &0.978  & 0.971 & 0.999 & 0.999 \\
         & k-hop MI similarity & 0.592 & 0.626 & 0.503 & 0.587 \\
         & ICC Similarity &0.935  & 0.791 & 0.936 &0.814  \\
         & Reliability Similarity & 0.823 & 0.772& 0.808 & 0.781 \\
         - Average & Between-table fidelity &0.901 & 0.863 & 0.896  & 0.877  \\
         \multicolumn{6}{l}{\textbf{Machine Learning efficacy}} \\   
         \multirow[t]{2}{*}{- Average} & Student-level efficacy & 0.963 & 0.937 & 0.972 & 0.894 \\
          & School-level efficacy & 0.825 & 0.837 & 0.948 & 0.902\\  
          & Overall efficacy &  0.940& 0.920 & 0.966 & 0.896\\              
        \bottomrule
    \end{tabular}
    \begin{tablenotes}
        \small
        \item NOTE: Scores represent similarity between the real and synthetic data, ranging from 0 to 1. Higher values denote greater similarity. The dashes represent that the metrics are not applicable to the given data type.
    \end{tablenotes}
    \end{threeparttable}
\end{table}

\newpage

\vspace{\fill}\clearpage

\section{Additional Simulations}

\subsection{Demonstration 1: AI-Based Simulations Using CTGAN-Generated Synthetic Data}\label{app:d1_CTGAN}

\begin{figure}[!ht]
    \centering
    \includegraphics[width=0.9\linewidth]{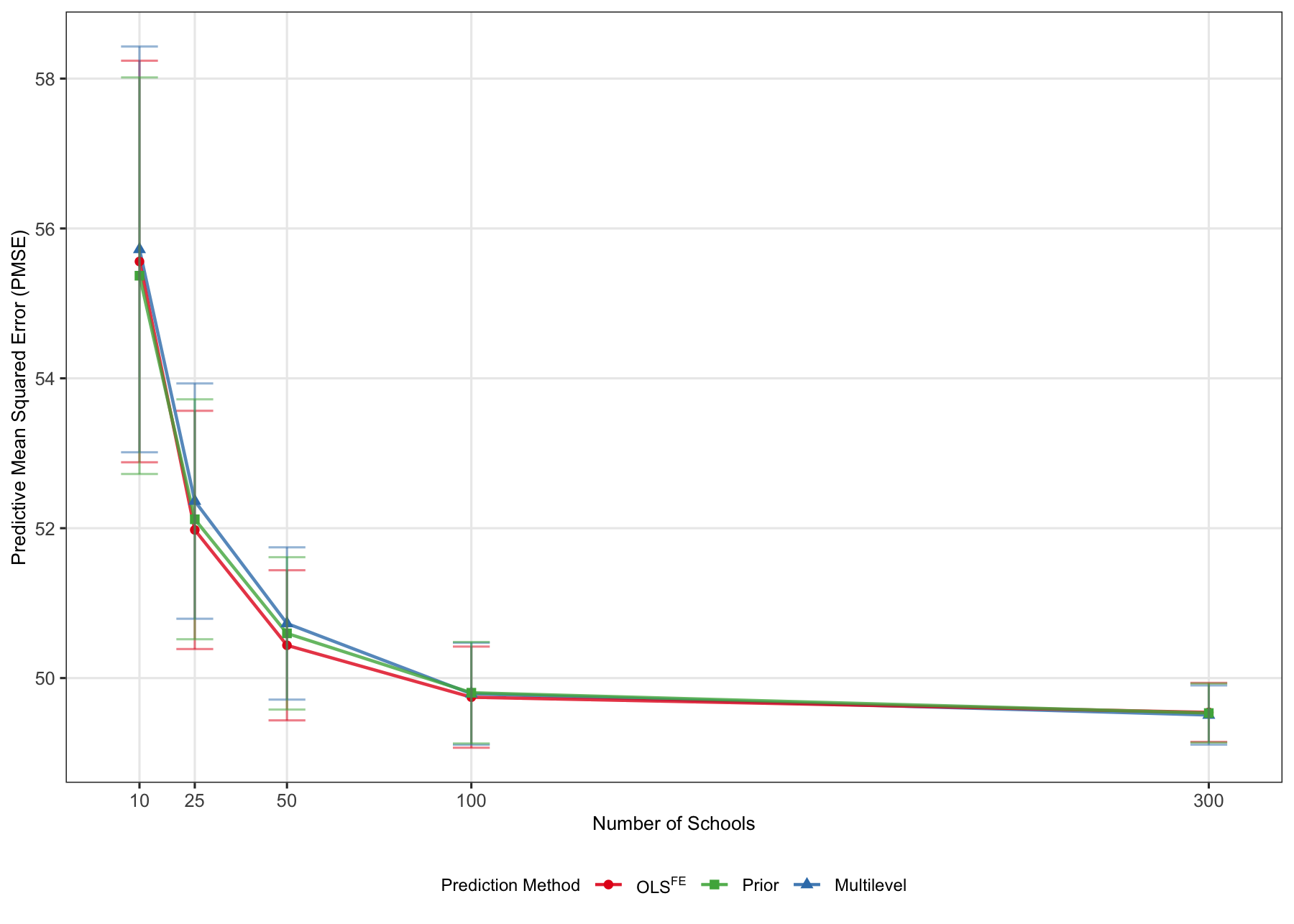}
    \caption{Results on the predicted mean squared error under the AI-based simulations using CTGAN-generated synthetic data.}
    \label{fig:ctgan_prediction}
\end{figure}

Although CTGAN achieved reasonably high overall similarity to the real data, its fidelity was weak in several aspects of between-table fidelity, notably ICC similarity and reliability similarity. Thus, the CTGAN-generated synthetic data did not pass our quality evaluation.  Nevertheless, for comparison purposes, we utilized the synthetic datasets generated by CTGAN in our AI-based simulation study. The simulation results are summarized in Figure~\ref{fig:ctgan_prediction}. We find no meaningful differences in performance among the three prediction methods. This finding strongly emphasizes that using insufficient-quality synthetic data undermines the accurate evaluation of how statistical methods perform in the real world.

\subsection{Demonstration 2: Traditional Simulations with Adjusted True Values}\label{app:demo2_huang_sim2}

\begin{figure}[!ht]
    \centering
    \includegraphics[width=\linewidth]{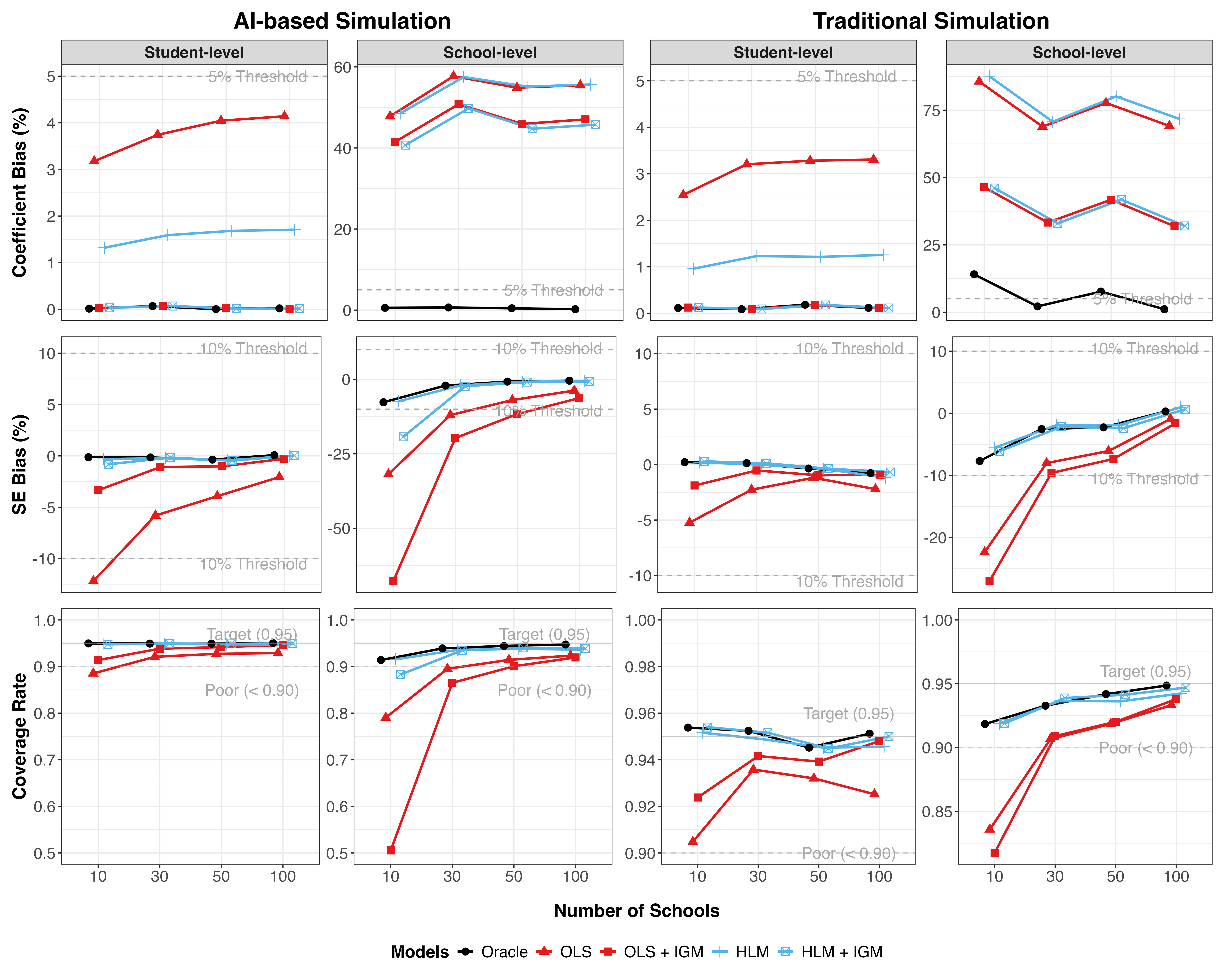}
    \caption{Results of the four models under the AI-based and adjusted traditional simulations. Jitter is added to clarify overlapping curves.}
    \label{fig:demo2_02}
\end{figure}

To increase the comparability between the AI-based and traditional simulations, we adjusted the true parameter values in the traditional simulation. Specifically, we changed the student-level coefficient from 5 to 0.223, and the school-level coefficient from 2 to -0.010. We also set the correlation between the school- and student-level variables to be the same as the empirical data. The results of this adjusted traditional simulation are summarized in Figure~\ref{fig:demo2_02}. Under this setting, we observe an increasing trend in coverage rate as the number of schools increases in both simulations. This confirms that the divergent coverage trends shown in Figure~\ref{fig:demo2} stem from the different scales of the true coefficients.

\newpage

\end{document}